\documentclass[9pt,twocolumn,twoside]{pnas-new}

\templatetype{pnasresearcharticle} 

\title{Internal States of Hadrons in Relativistic Reference Frames}

\author[a]{N.O.~CHUDAK}
\author[a]{K.K.~MERKOTAN}
\author[a]{D.A.~PTASHYNSKYY}
\author[a]{O.S.~POTIYENKO}
\author[b,d]{M.A.~DELIYERGIYEV}
\author[c,d]{A.V.~TYKHONOV}
\author[a,d]{G.O.~SOKHRANNYI}

\author[e]{O.V.~ZHAROVA}
\author[a]{O.D.~BEREZOVS’KYI}
\author[a]{V.V.~VOITENKO}
\author[a]{YU.V.~VOLKOTRUB}
\author[a]{I.V.~SHARPH}
\author[a]{V.D.~RUSOV}

\affil[a]{Department of Theoretical and Experimental Nuclear Physics, Odessa National Polytechnic University (Shevchenko av. 1, Odessa, 65044, Ukraine)}

\affil[b]{Department of High Energy Nuclear Physics, Institute of Modern Physics (Nanchang Road 509, 730000 Lanzhou, China)}

\affil[c]{D\'{r}partement de physique nucl\'{e}aire et corpusculaire, Universit\'{e} de Gen\'{e}ve (CH-1211 Geneva 4, Switzerland)}%

\affil[d]{Department of Experimental Particle Physics, Jo\v{z}ef Stefan Institute 
	(Jamova 39, SI-1000 Ljubljana, Slovenia)}%

\affil[e]{Department of Higher Mathematics and Modeling Systems, Odessa National Polytechnic University (Shevchenko av. 1, Odessa, 65044, Ukraine)}

\leadauthor{N.O.~CHUDAK} 

\correspondingauthor{\textsuperscript{2}Contact e-mail: siiis$\textcircled{a}$te.net.ua}

\keywords{hadrons $|$ reference frame $|$ bound states of quarks $|$ hadron scattering $|$ state transformation}

\begin{abstract}

The internal state of a composite particle and its transformation, when
changing from the reference frame, where this composite particle is at rest, to a reference frame,
where it moves relativistically, have been considered. 
It is supposed that the internal state of the composite particle in its rest frame can be considered in the non-relativistic approximation. This internal state is shown to remain the same, when changing from one inertial reference frame to another one. In other words, a particle that is spherically symmetric in its rest frame does not change its form in any other reference frame and does not undergo the Lorentz contraction in the direction of motion of any reference frame with respect to the rest one. A possible application of the results obtained to describe the scattering of hadrons considered as bound states of quarks has been discussed.

\end{abstract}

\doi{Submitted to Ukr.J.Phys. ~~\url{http://ujphys.bitp.kiev.ua}}

\begin{document}

\verticaladjustment{-2pt}

\maketitle
\thispagestyle{firststyle}
\ifthenelse{\boolean{shortarticle}}{\ifthenelse{\boolean{singlecolumn}}{\abscontentformatted}{\abscontent}}{}


\section{Introduction}
\dropcap{I}n our previous work Ref.\cite{ujp}, it was shown that the processes of elastic hadron scattering can be described in the framework of the Laplace method. However, the corresponding calculations were carried out only for model scalar theories, which allowed experimental results to be reproduced only at a qualitative level Ref.\cite{cej,Sharph:2015eka}. On the other hand, the obtained qualitative agreement gave hope for that the key features in the behavior of experimentally observed quantities can be described in the framework of perturbation theory even in the case of strong interaction. Therefore, there emerged an idea to apply the Laplace method in the framework of quantum chromodynamics perturbation theory Ref.\cite{Sharf:2012vy}.

Following this way, we are faced with the known problem: in diagrams, the quark and gluon lines are used, while we deal in the initial and final states with the bound states of quarks -- the hadrons. As a result and contrary to what takes place in the “standard” scattering theory, the interaction between quarks cannot be switched-on or -off. Accordingly, neither the state nor the Hamiltonian for the system of scattered particles asymptotically approach the corresponding quantities for free quarks. Therefore, we obtain two problems: for the state and for the Hamiltonian. The former consists in that how the initial and final scattering states can be assigned with regard for the interaction between quarks. The essence of the latter is that, while considering the scattering amplitude with the use of the diagram technique considerably related to the switching-on and -off of the interaction, we obtain an energy-momentum conservation law, which is applied to the four-momenta of quarks, rather than hadrons, i.e. contrary to what it occurs in experiment. In the present paper, we examine the problem of states. The solution of the problem concerning the “correct” form for the energy-momentum conservation law was considered in our works Ref.\cite{Volkotrub:2016, Chudak:2016}. The results obtained in those works were substantially based on the conclusions presented in this paper below.

As a rule, hadrons in scattering processes are described in the framework of the parton model Ref.\cite{Feynman:1972}. However, the relativistic description of the internal hadron state demands that a considerable number of many-parton distribution functions should be assigned, if we do not confine ourselves to the inclusive description, and this task substantially complicates the problem Ref.\cite{Diehl:2011yj, Strikman:2011zz}. The solution of this problem can be made easier, by using the following speculations. If we assume that a free hadron consists of a certain number of certain constituent quarks in its initial or final scattering state, this means that new constituent quarks cannot be born as a result of the interaction between those quarks. As a result, we suppose that at least some effects of elastic and inelastic hadron scatterings can be described, if the internal state of a free hadron is considered in the rest frame of this hadron, rather than in the non-relativistic approximation. This approach does not exclude the fact that the purely relativistic description should be applied to some specific effects (see, e.g., Ref.\cite{Kobushkin:1977}).

It should be emphasized that the matter concerns just a free hadron, before or after its scattering. In the course of the scattering, the interaction between the quarks belonging to different hadrons must undoubtedly be described relativistically. Such a description is not a subject of this work, but it was made in Ref.\cite{Volkotrub:2016, Chudak:2016}. Nevertheless, the initial and final states at the scattering process contain several hadrons each. Therefore, generally speaking, we cannot choose the reference frame in such a way that it would be a rest frame for all those hadrons, or at least that all hadrons in this reference frame would be non-relativistic.

Hence, there emerges a problem to transform the non-relativistic internal state and the Hamiltonian, when changing from particle’s rest frame to a reference frame, where this particle moves relativistically. The essence of this problem can be explained, by using the following simple example. Suppose that there are the simplest quantum-mechanical non-relativistic system – a hydrogen atom – with the spherically symmetric ground state and an inertial observer moving relative to this atom with a relativistic velocity. We would like to know the coordinates and momenta of the particles composing this system, as measured by this observer. More specifically, which probability amplitude describes the results of measurements, and how is this amplitude related to the probability amplitude measured in the rest system of a hydrogen atom (the rest frame for the center-of-mass of the constituting particles)?

If the hydrogen atom could be considered from the standpoint of classical rather than quantum mechanics, the procedure could be as follows. First, a standard problem of two bodies interacting by means of a given potential is solved in atom’s rest frame. Then we can apply the Lorentz transformations to this solution, rather that consider a relativistic problem on the self-consistent dynamics of three interacting objects: the nucleus, electron, and electromagnetic field (the dynamic characteristics of each of them are not fixed in this case and should be determined in the course of solution). Doing in such a manner, we could avoid a necessity to apply the relativistic description of the interaction field between the nucleus and the electron. Our purpose in this work is to implement an approach of this kind, but in the framework of quantum rather than classical mechanics. Namely, we intend to use the Schr\"{o}dinger equation with a definite potential energy in the reference frame of the center-of-mass of a two-particle system, to transform the state determined in the rest frame into the reference frame that moves relativistically with respect to the center-of-mass reference frame, and to avoid the quantum-mechanical field description associated with the creation and destruction of virtual particles in the system.

Bearing in mind that we are interested in hadrons, we will not consider the hydrogen atom below, but a meson consisting of a quark and an antiquark as the example of a two-particle system. Afterward, we will apply the obtained results to more complicated three-quark systems, baryons, assuming that the internal states of those particles, similarly to a hydrogen atom, can be described in the non-relativistic approximation in their rest frames. The example with a hydrogen atom was given to emphasize that, in this work, we do not discuss a possibility to apply the non-relativistic approximation in the rest frame of the combined particle, as we have no necessity to do this in the case of a hydrogen atom. We consider the problem in the following formulation: supposing that the internal state of a combined particle is non-relativistic in its rest frame, we have to determine its state in a reference frame that moves at a relativistic velocity relative to the rest reference frame.

The outlined problem is rather non-typical. As a rule, various quantities associated with the same event are measured in different reference frames. But in the case of the probability amplitude for many-particle systems, the situation is different. Really, consider two inertial observers, which we call \textit{unprimed} and \textit{primed} respectively. 
From the viewpoint of the unprimed observer, the probability amplitude for a two-particle system  (it will be denoted as $\Psi \left( t,{{{\mathbf{r}}}_{1}},{{{\mathbf{r}}}_{2}} \right)$)  describes the result of coordinate measurements that were performed simultaneously in the unprimed system and at the time moment ${t}$ according to the clock in this system.  Analogously, the probability amplitude ${\Psi }'\left( {t}',{{{{\mathbf{r}}'}}_{1}},{{{{\mathbf{r}}'}}_{2}} \right)$ for the primed observer describes the result of coordinate measurements performed simultaneously with respect to this observer, at the time moment ${t}'$ according to his/her clock.

This is an essential difference between the problem concerned and the classical problem about the Lorentz contraction. In the latter, the coordinate measurement of rod’s ends must be simultaneous in the reference frame, relative to which the rod moves. However, it can be non-simultaneous in rod’s rest frame. Therefore, rod’s length can be calculated in terms of coordinates of the same events, but measured in different reference frames. In our case, a pair of events consisting in that one observer detects particles in close vicinities of some points and a similar pair of events for the other observer comprise substantially different pairs of events. It is so because the events in the first pair must be simultaneous for the first observer, as well as two events in the second pair for the second observer. Therefore, those two observers cannot use the same measurement, when expressing its results with the use of the variables of the corresponding own reference frame. Each of the observers should realize his/her own, simultaneous with respect to his/her, independent measurement. As a result, there is no relation between the $ \left( t,{{{\mathbf{r}}}_{1}},{{{\mathbf{r}}}_{2}} \right)$ --
and $\left( {t}',{{{{\mathbf{r}}'}}_{1}},{{{{\mathbf{r}}'}}_{2}} \right)$ --values, because such a relation can exist only between the time coordinates of the same event measured in different reference frames. In other words, there are no relations, like Lorentz transformations, between the arguments of the probability amplitudes in both reference frames. (Hereafter, the word argument has a sense of the variable, on which a function depends, rather than an $argument$ of the function value as a complex number.) Therefore, the conclusions about the length contraction or the time dilation as a consequence of Lorentz transformations become invalid.

The conclusion that neither Lorentz nor any other transformations can relate arguments of many-particle probability amplitudes in different reference frames considerably distinguishes the approach of this work from approaches applied in other works on this subject, which are known to us. The authors of those works explicitly or implicitly assume that the arguments of many-particle probability amplitudes in different reference frames can be related to one another by Lorentz transformations. In particular, the indicated problem concerning the simultaneity was already considered in the literature. For instance, in Ref.\cite{FAUSTOV1973176}, the simultaneity that is invariant with respect to Lorentz transformations was proposed to be defined as a simultaneity in the center-of-mass frame. In the most known work on this subject Ref.\cite{PhysRev.84.1232}, an analog of two-particle probability amplitudes was introduced as a matrix element of the product of two one-particle creation operators in the Heisenberg representation. The arguments of those two operator functions were considered as four-vectors with respect to Lorentz transformations, which brings us to the well-known problem of relative times. A further projection of the Bethe–Salpeter function on a certain space-like hypersurface in the Minkowski space, which is used in the quasipotential method Ref.\cite{Logunov1963}, was aimed at avoiding the problem of relative times by introducing an invariant time-like variable; i.e. this procedure also assumes that the arguments of probability amplitudes can be interrelated by means of Lorentz transformations.

The same is true for the works, in which the dynamics of a light front was considered (see, e.g., Refs.\cite{Brodsky:1997de, Heinzl:1998kz, Terentev:1976}). In the seminal work Ref.\cite{RevModPhys.21.392} on this subject, the main purpose was to reject the idea of simultaneous description; however, the matter concerned the construction of expressions for the generators of the Poincare group. At the same time, the analysis of the front-form dynamics from the viewpoint of state space Ref.\cite{Brodsky:1997de, Heinzl:1998kz} in which those generators operate again brings about “light-cone wave functions”, which arguments are also assumed to be related by means of Lorentz transformations.

The attention should be attracted to the fact that the speculations given above about the impossibility to interrelate the arguments of many-particle probability amplitude have nothing to do with our intention to apply the non-relativistic approximation in one of the reference frames. In the relativistic situation, when the state is described by a Fock column Ref.\cite{Bogolubov:1959,Berezin:1966}, the above-mentioned problem of simultaneous measurement arises for every component of this column starting from the second one. Therefore, on the basis of the same reasons that were presented above, we again arrive at a conclusion that the arguments of the components in the Fock columns, which describe the same state of a relativistic quantum system in different reference frames, cannot be related to one another in any way. A key method for the solution of the problem concerning the Fock state transformation, when changing from one inertial reference frame to another one, is given by the field quantization postulate formulated in work Ref.\cite{Bogolubov:1959}. According to it, the role of the generators of a Lorenz group representation in the Fock space is played by the components of the angular momentum operator for the corresponding relativistic system. More specifically, in our situation, this means the following.

It is known that the change from one inertial frame to another one can be presented as a product of two rotations and one boost. The indicated problem of simultaneity evidently does not arise in the case of rotations, so that the problem of state transformation does not arise as well. Therefore, only the boost case will be considered below.

Taking into account that it is enough to consider a boost only along one of coordinate axes, let us analyze the case of the boost along the axis $OZ$. The boost rapidity will be designated as $Y$. According to the quantization postulate Ref.\cite{Bogolubov:1959}, the generator of the state transformation in this case is the operator of the angular momentum component $\hat{M}_{03}$. That is the state $\left| \Psi^{\prime} \right\rangle$ (the non-relativistic probability amplitude
or the relativistic Fock column) in the primed reference frame is related to the corresponding state $\left| \Psi \right\rangle$ in the unprimed reference frame by means of the relation
\begin{equation}
\begin{split}
&\left| \Psi^{\prime} \right\rangle =\hat{U}(Y)\left| \Psi \right\rangle,\\
&\hat{U}(Y) =\exp \left( i{{{\hat{M}}}_{03}}Y \right).\\
\end{split}
\label{eq:Fokstan}
\end{equation}

The generator $\hat{M}_{03}$ is an operator-valued functional of the creation and annihilation operators in the Fock space. If the field operators and the Fock state $\left| \Psi \right\rangle$ are considered in the Heisenberg representation, then, in accordance with Noether’s theorem, the generator $\hat{M}_{03}$ is time-independent. Therefore, in any other representation, where both the state and the generator depend on the time, we obtain that the time value in the expressions for the state and the generator is the same. Furthermore, $\hat{M}_{03}$ is an integral of the corresponding density over the coordinates. Hence, the action of this operator in \eqref{eq:Fokstan} does not result in the appearance of new independent variables. Therefore, since the variables, on which the components of the column $\left| \Psi \right\rangle$ depend, are not expressed in any way, we obtain that the components of the column $\left| \Psi^{\prime} \right\rangle$ depend on the same variables. So, the action of the operator $\hat{U}(Y)$ changes only the form of dependences. Hence, being unable to establish a relationship between the probability amplitude values corresponding to the same event, when considering the problem of a probability amplitude transformation between inertial reference frames, we can establish, nevertheless, a relationship between those values obtained at the same values of arguments, as is done, when considering internal symmetries. More specifically, if we consider \eqref{eq:Fokstan}, e.g., in the coordinate representation, its left-hand side contains the time and coordinates that correspond to some events, which are simultaneous in the initial reference frame. At the same time, on the right-hand side of this equality, other events are considered, and their spatial and temporal coordinates are the same as on the left-hand side, but expressed in the new reference frame and simultaneous with respect to it. Proceeding from the aforesaid, the notation $\Psi^{\prime}\left( {t}',{{{{\mathbf{r}}'}}_{1}},{{{{\mathbf{r}}'}}_{2}} \right)$  will be used for the two-particle probability amplitude in the primed coordinate system.

In the relativistic case, in addition to the difficulties indicated above and associated with the definition of a Fock state, we face another one, which consists in that the operator $\hat{M}_{03}$ for systems with interaction does not depend quadratically on the creation and annihilation operators. Therefore, the functional integral that describes its action in the Fock space Ref.\cite{Berezin:1966} is not Gaussian. In other words, even if we could define a Fock state, the problem of its transformation with the use of \eqref{eq:Fokstan} would be very difficult. This is another argument for the attempt to simplify the situation within the non-relativistic approximation.

However, in this case, we should construct a corresponding non-relativistic approximation for the generator $\hat{M}_{03}$. If such an approximation is constructed in the center-of-mass frame for a bound system, the operator $\hat{M}^{\prime}_{03}$ can be expressed in any other reference frame in terms of $\hat{M}_{03}$ and other angular momentum components, which can also be defined in the non-relativistic approximation with the help of the tensor transformation rule. The following section is devoted to the formulation of such approximations.

Before proceeding to the solution of the described problem, we should pay attention to the following capability of its simplification. Let us consider a hadron in its rest frame. Here, the state of the system has to be an eigenstate for the operator of total momentum of all constituting particles, $\hat{\mathbf{P}}$, and to correspond to the zero eigenvalue. Even before making the non-relativistic approximation, the temporal evolution of the Fock state $\left| \Psi \right\rangle$ in the system of particles that form the hadron can be written in the form
\begin{equation}
\left| \Psi \left( t \right) \right\rangle =\exp \left( -i\hat{H}t \right)\left| \Psi \left( t=0 \right) \right\rangle ,
\end{equation}
where $\hat{H}$ is the relativistic Hamiltonian of the system of fields, whose quants are making up the hadron. According to Ref.\cite{Bogolubov:1959}, in the reference frame obtained from the initial one by applying the boost transformation, we have
\begin{equation}
\left| {\Psi }'\left( t \right) \right\rangle =\hat{U}\left( Y \right)\left( \exp \left( -i\hat{H}t \right)\left| \Psi \left( t=0 \right) \right\rangle  \right).
\label{eq:peretvorenna}
\end{equation}
Here, $\hat{U}\left( Y \right)$ is the unitary operator of state transformation owing to the boost with rapidity $Y$, which is defined by relation \eqref{eq:Fokstan}. Taking into account that we consider an eigenstate of the total momentum corresponding to the zero eigenvalue, relation \eqref{eq:peretvorenna} can be rewritten in the form
\begin{eqnarray}
& \left| {\Psi }'\left( t \right) \right\rangle =\hat{U}\left( Y \right)\left( e^{ -i\left( \hat{H}t-\left( \hat{\mathbf{P}}\cdot \hat{\mathbf{R}} \right) \right) } \left| \Psi \left( t=0 \right) \right\rangle  \right),
\label{eq:peretvorenna1}	
\end{eqnarray}
where $\mathbf{R}$ is a set of three arbitrary coordinates. Their specific choice is not important, because operator \eqref{eq:peretvorenna1} acts on that eigenfunction of the operator $\hat{\mathbf{P}}$, which corresponds to its zero eigenvalue. However, we may consider now the set of four numbers ($t$ and $\mathbf{R}$) as components of a four-vector with respect to the Lorentz transformations. The set of the operators $\hat{H}$ and $\hat{\mathbf{P}}$ can also be considered as an operator four-vector. This circumstance can be used as follows

Let us rewrite expression \eqref{eq:peretvorenna1} in the form
\begin{eqnarray}
& \left| {\Psi }'\left( t \right) \right\rangle =\hat{U}\left( Y \right)\hat{u}\left( x \right){{{\hat{U}}}^{-1}}\left( Y \right)\hat{U}\left( Y \right)\left| \Psi \left( t=0 \right) \right\rangle,
\label{eq:peretvorenna2}	
\end{eqnarray}
where we use the following notations:
\begin{equation}
\begin{split}
x&\equiv \left( t,{{R}_{x}},{{R}_{y}},{{R}_{z}} \right),\\
\hat{u}\left( x \right)&\equiv e^{ -i\left( \hat{H}t-\left( \hat{\mathbf{P}}\cdot \hat{\mathbf{R}} \right) \right)},
\end{split}
\label{eq:poznachenna_x_u}	
\end{equation}
were introduced. The expression $\hat{U}\left( Y \right)\hat{u}\left( x \right){{{\hat{U}}}^{-1}}$ is formally identical to that arising at the transformation of operator field functions Ref.\cite{Bogolubov:1959}. Therefore, designating the matrix of a boost along the axis $OZ$ as ${{\Lambda }^{\left( 0 \right)}}\left( Y \right)$, we obtain
\begin{eqnarray}
\hat{U}\left( Y \right)\hat{u}\left( x \right){{\hat{U}}^{-1}}\left( Y \right)=\hat{u}\left( {{\Lambda }^{\left( 0 \right)}}\left( Y \right)x \right).
\label{eq:poznachenna_u1}	
\end{eqnarray}
Then instead of \eqref{eq:peretvorenna2}, we can write:
\begin{equation}
\begin{split}
\left| {\Psi }'\left( t \right) \right\rangle &=\exp \left( -it\left( \operatorname{ch}\left( Y \right)\hat{H}+\operatorname{sh}\left( Y \right){{{\hat{P}}}_{z}} \right) \right)\\
&\times\exp \left( i{{R}_{z}}\left( \operatorname{sh}\left( Y \right)\hat{H}+\operatorname{ch}\left( Y \right){{{\hat{P}}}_{z}} \right) \right)\\
& \times \exp \left( i\left( {{R}_{x}}{{{\hat{P}}}_{x}}+{{R}_{y}}{{{\hat{P}}}_{y}} \right) \right)\hat{U}\left( Y \right)\left| \Psi \left( t=0 \right) \right\rangle.
\end{split}
\label{eq:peretvorenna3}	
\end{equation}

Till now, the relativistic operators of energy and momentum were considered as $\hat{H}$ and $\hat{\mathbf{P}}$, respectively. However, they are related to the initial reference frame, where, according to the considered problem, the non-relativistic approximation can be applied. In this approximation, those operators can be substituted by the non-relativistic internal Hamiltonian for the system of quarks composing the hadron, and the non-relativistic operator of momentum of this system. In this non-relativistic approximation, the quantity $\left| \Psi \left( t=0 \right) \right\rangle$ can be substituted by the coordinate part of the probability amplitude for the energy eigenstate of a two-particle (quark and antiquark) system. In addition, in the limiting case of low rapidities $Y$, we see that the coordinates of the center of mass must be selected as arbitrary coordinates of vector ${\mathbf{R}}$:
\begin{align}
\mathbf{R}=\mathbf{R}\left( {{{\mathbf{r}}}_{1}},{{{\mathbf{r}}}_{2}} \right)=\frac{{{m}_{1}}{{{\mathbf{r}}}_{1}}+{{m}_{2}}{{{\mathbf{r}}}_{2}}}{{{m}_{1}}+{{m}_{2}}}.
\label{eq:centr_mas}
\end{align}

As will be shown latter, if one takes, for $\left| \Psi \left( t=0 \right) \right\rangle$, the energy eigenstate of a two-particle bound system that corresponds to the smallest eigenvalue, this state remains invariant after the action of the operator $\hat{U}\left( Y \right)$:
\begin{equation}
\hat{U}\left( Y \right)\left| \Psi \left( t=0 \right) \right\rangle=\left| \Psi \left( t=0 \right) \right\rangle,
\label{eq:U_operator}
\end{equation}
In addition, bearing in mind that $\hat{H}$ is the Hamiltonian of the system of bound particles in their center-of-mass frame, and $\left| \Psi \left( t=0 \right) \right\rangle$ is its eigenstate corresponding to its smallest eigenvalue, we have
\begin{equation}
\hat{H}\left( Y \right)\left| \Psi \left( t=0 \right) \right\rangle=m_{\mu}\left| \Psi \left( t=0 \right) \right\rangle.
\label{eq:Hamiltonia_operator}
\end{equation}
From \eqref{eq:peretvorenna3}, taking all that and the reasons given above into account, we obtain a “correct” dependence on the time and the center-of-mass coordinates in the new reference frame:
\begin{equation}
\left| \Psi^{\prime} \right\rangle=\exp\left(-i \left(\sqrt{m_{\mu}^{2}+\mathbf{P}^{2}}t-\left(\mathbf{R}(\mathbf{r}_{1},\mathbf{r}_{2})\cdot\mathbf{P}\right)  \right) \right),
\label{eq:Psi_prime_framDepen}
\end{equation}
where $\mathbf{P}$ is the momentum of a bound particle in the considered reference frame, and $\mathbf{R}(\mathbf{r}_{1},\mathbf{r}_{2})$ is expressed by \eqref{eq:centr_mas}.

We would like to attract attention to that, when changing from the center-of-mass frame to another inertial one, this “correct” dependence appeared not due to a transformation of the coordinates and the time, as this occurs for an ordinary plane wave, but exclusively due to the form transformation of dependence \eqref{eq:poznachenna_u1} on the same variables, as was discussed above after formula \eqref{eq:Fokstan}. Changing from the center-of-mass frame to various other inertial reference frames, dependence \eqref{eq:Psi_prime_framDepen} will be obtained in each of them, and, as was discussed above, $\mathbf{r}_{1}$ and $\mathbf{r}_{2}$ will be the coordinates of particles that are measured simultaneously in the corresponding reference frame. Hence, we obtain the same dependence of the state on the variables in different inertial reference frames, as the relativity principle demands.

Thus, the discussed simplification, which is reached by applying Eqs.(\ref{eq:peretvorenna1})--(\ref{eq:peretvorenna3}), consists in that we have no need to describe the transformation of the whole probability amplitude for the energy eigenstate, when changing from the quark-antiquark center-of-mass reference frame to another one. We may confine ourselves to the transformation of only the coordinate part of this probability amplitude. Hence, the further consideration concerns two issues: (i) How can a non-relativistic approximation for the generator $\hat{M}_{03}$ be constructed? and (ii) How can the operator $\exp \left( i{{{\hat{M}}}_{03}}Y \right)$ be applied to the coordinate part of hadron's internal state in hadron's rest frame?

At the end of Introduction, we would like to distinctly emphasize those approximations that are used in this work. Note that we do not deal with the relativistic theory of bound states. We consider a problem, in which the internal state of a bound particle is given in the reference frame of particle’s center-of-mass, and it is non-relativistic. We should determine this state in a reference frame that moves at a relativistic velocity with respect to the center-of-mass reference frame. We hope for that, in the framework of this approximation, it will be possible to describe the main properties of relativistic elastic and inelastic hadron scatterings Ref.\cite{Volkotrub:2016, Chudak:2016}. In the rest frame of the hadron, its internal state is described by a two-particle probability amplitude (this is a solution of the Schr\"{o}dinger equation), and hadron’s mass is the smallest eigenvalue of the corresponding non-relativistic Hamiltonian Ref.\cite{Volkotrub:2016, Chudak:2016}. The non-relativistic approximation is also used for the component $\hat{M}_{03}$ of the angular momentum tensor in the rest frame of a bound particle.

\section{Approximation of Lorentz Transformation Generators by Differential Operators}

The component $\hat{M}_{03}$ of the angular momentum tensor in terms of differential operators looks like
\begin{align}
{{\hat{M}}_{03}}=i\left( t\frac{\partial }{\partial z}+z\frac{\partial }{\partial t} \right).
\label{eq:dif1particl}
\end{align}
Note that the expression of the generators in terms of differential operators can be obtained by considering a certain function of the coordinates and the time and by making the relevant substitution of independent variables in this function. However, as was

marked in Introduction, the change of independent variables is impossible in our case. Therefore, relation \eqref{eq:dif1particl} can be understood only as a limit for the “correct” relativistic operator $\hat{M}_{03}$, when performing the non-relativistic approximation. Then, the following question arises: To what limit does this operator tend in the case of a many-particle system? Taking into account that the spatial components of the angular momentum are expressed as sums of corresponding one-particle operators, we may assume that the components, for which one of the subscripts equals zero, are additive as well. Then, for a two-particle system, we have
\begin{align}
{{\hat{M}}_{03}}=i\left( t\left( \frac{\partial }{\partial {{z}_{1}}}+\frac{\partial }{\partial {{z}_{2}}} \right)+\left( {{z}_{1}}+{{z}_{2}} \right)\frac{\partial }{\partial t} \right).
\label{eq:pripucenna}
\end{align}
As already noted, when making the non-relativistic approximation, the quantity  $\left| \Psi \left( t=0 \right) \right\rangle$,  in \eqref{eq:peretvorenna3}, can be replaced, in our case, by coordinate part  $\psi \left( {{{\mathbf{r}}}_{1}},{{{\mathbf{r}}}_{2}} \right)$. 
of the energy eigenstate. This function does not depend on the time, and it is the eigenfunction of the operator of total system momentum corresponding to the zero eigenvalue.

Taking into account that operator \eqref{eq:pripucenna} can be written in the form
\begin{align}
{\hat M_{03}} =  - t{\hat P_z} + \left( {{z_1} + {z_2}} \right)i\frac{\partial }{{\partial t}},
\label{eq:pripucenna1}
\end{align}
we arrive at a conclusion that the function $\psi \left( {{{\mathbf{r}}}_{1}},{{{\mathbf{r}}}_{2}} \right)$ is also an eigenfunction of the operator $\hat{M}_{03}$ and corresponds to its zero eigenvalue.

This fact can also be explained by the following reasons. Since the quark-antiquark center-of-mass reference frame is the initial one, we have
\begin{align}
\psi \left( {{{\mathbf{r}}}_{1}},{{{\mathbf{r}}}_{2}} \right)=\psi \left( {{{\mathbf{r}}}_{2}}-{{{\mathbf{r}}}_{1}} \right).
\label{eq:cm_koord_chast}
\end{align}
If the variables $\mathbf{r}_{1}$ and $\mathbf{r}_{2}$ in the expression
\begin{align}
i\left( t\left( \frac{\partial }{\partial {{z}_{1}}}+\frac{\partial }{\partial {{z}_{2}}} \right)+\left( {{z}_{1}}+{{z}_{2}} \right)\frac{\partial }{\partial t} \right)\psi \left( {{{\mathbf{r}}}_{2}}-{{{\mathbf{r}}}_{1}} \right),
\label{eq:M03psi}
\end{align}
are replaced by new ones,
\begin{align}
{\mathbf r_ + } = {\mathbf r_1} + {\mathbf r_2},~~~{\mathbf r_ - } = {\mathbf{r_1}} - {\mathbf{r_2}},
\label{eq:novi_rplus_rminus}
\end{align}
then the operator in \eqref{eq:M03psi} will depend only on $z$ component of vector $\mathbf{r_ {+}}$ as ${{z}_{+}}$, and function on which this operator acts will depend only on $z$ component of vector $\mathbf{r_ {-}}$ as ${{z}_{-}}$. Therefore, from the reasons given above, a conclusion can be drawn that
\begin{align}
\exp \left( i{{{\hat{M}}}_{03}}Y \right)\psi \left( {{{\mathbf{r}}}_{2}}-{{{\mathbf{r}}}_{1}} \right)=\psi \left( {{{\mathbf{r}}}_{2}}-{{{\mathbf{r}}}_{1}} \right).
\label{eq:visnovok}
\end{align}
i.e. meson's internal state does not vary when changing to a new reference frame.

In all previous discussions, the corresponding component of the orbital angular momentum tensor was considered as the generator of $\hat{M}_{03}$. Note that, in the case of free bispinor field, one can see from the explicit expression for the spin contribution to the angular momentum tensor Ref.\cite{Bogolubov:1959} that the contributions of those tensor components, for which at least one of their subscripts equals zero, vanish.  The operator of interaction between the bispinor and gauge fields does not contain derivatives of the bispinor field components and, therefore, does not contribute to the tensor of spin angular momentum. Therefore, the spin contribution to the ``correct'' relativistic operator $\hat{M}_{03}$ equals zero. This means that only the orbital contribution to $\hat{M}_{03}$ can be considered in the non-relativistic limit.

From whence, we may draw conclusion that all the reasoning above can be applied not only to mesons, but also to baryons, because the presence of their nonzero spin changes nothing. Assuming the additivity of all components of the angular momentum, we obtain that, for a baryon, instead of \eqref{eq:pripucenna}, we have
\begin{equation}
{{{\hat{M}}}_{03}}=i\left( t\left( \frac{\partial }{\partial {{z}_{1}}}+\frac{\partial }{\partial {{z}_{2}}}+\frac{\partial }{\partial {{z}_{3}}} \right) +\left( {{z}_{1}}+{{z}_{2}}+{{z}_{3}} \right)\frac{\partial }{\partial t} \right).
\label{eq:pripucenna3}
\end{equation}
This operator is also expressed in terms of the operator of the $z$-component of the total momentum of the system. Therefore, the action of this operator on the eigenfunction of the total momentum operator corresponding to the zero eigenvalue also gives zero.

The speculations in this section possess two essential
shortcomings. First, the “correct” relativistic operator
$\hat{M}_{03}$ is not realized in terms of differential operators, but is given in the second-quantization representation. That is why it is reasonable to seek its non-relativistic limit in this representation. In addition, we substantially used assumptions \eqref{eq:pripucenna} and \eqref{eq:pripucenna3}. In the second-quantization representation, since the operator expressions do not depend on whether the operators are defined in the one- or many-particle space, those assumptions turn out unnecessary. Therefore, the considerations of this section can be regarded only as auxiliary. However, in the next section, we will demonstrate that the consideration of the problem in the second-quantization representation brings about the same result.

\section{Approximation of Lorentz Transformation Generators in the Second Quantization Representation}

Let the notation ${{\hat{q}}^{+}}\left( f,\nu ,c,\mathbf{r} \right)$ designate the non-relativistic creation quark operator in the coordinate representation of the second quantization. The indices $f,\nu,c$ describe the flavor, spin and color, respectively, of a quark created in the state that is characteristic of the radius-vector operator and corresponds to the eigenvalue $\mathbf{r}$.
The creation antiquark operator in the same state denoted as ${{\hat{\bar{q}}}^{+}}\left( f,\nu ,c,\mathbf{r} \right)$.
The corresponding annihilation operators are ${{\hat{q}}^{-}}\left( f,\nu ,c,\mathbf{r} \right)$ and ${{\hat{\bar{q}}}^{-}}\left( f,\nu ,c,\mathbf{r} \right)$. Then the coordinate part of meson’s internal state (the meson is regarded as a quark-antiquark system) can be presented in the form
\begin{equation}
\begin{split}
\left| \mu  \right\rangle =\int{d{{{\mathbf{r}}}_{2}}d{{{\mathbf{r}}}_{1}}\psi \left( \left| {{{\mathbf{r}}}_{2}}-{{{\mathbf{r}}}_{1}} \right| \right)}s\left( {{\nu }_{1}},{{\nu }_{2}} \right)c\left( {{c}_{1}},{{c}_{2}} \right)a\left( {{f}_{1}},{{f}_{2}} \right)  \\
\times {{{\hat{q}}}^{+}}\left( {{f}_{1}},{{\nu }_{1}},{{c}_{1}},{{{\mathbf{r}}}_{1}} \right){{{\hat{\bar{q}}}}^{+}}\left( {{f}_{2}},{{\nu }_{2}},{{c}_{2}},{{{\mathbf{r}}}_{2}} \right)\left| 0 \right\rangle .
\end{split}
\label{eq:coordinatna_castina}
\end{equation}
In this formula $s\left( {{\nu }_{1}},{{\nu }_{2}} \right)c\left( {{c}_{1}},{{c}_{2}} \right)a\left( {{f}_{1}},{{f}_{2}} \right)$ 
stand for the spin, color, and flavor, respectively, parts of the probability amplitude, whereas the function $\psi \left( \left| {{{\mathbf{r}}}_{2}}-{{{\mathbf{r}}}_{1}} \right| \right)$ describes the coordinate dependence of a probability amplitude in the quarkantiquark center-of-mass reference frame. Since we consider the coordinate part of the energy eigenstate, an eigenfunction of the non-relativistic Hamiltonian for the quark-antiquark system has to be taken as $\psi \left( \left| {{{\mathbf{r}}}_{2}}-{{{\mathbf{r}}}_{1}} \right| \right)$. As usual, the summation over the repeated indices is implied. We also use the standard notation $\left| 0 \right\rangle$ for the vacuum state.

Taking into account that the dependence of all quantities on the internal indices is insignificant for the issues considered in this section, let us use the single-letter notation $\xi$ for the set of indices $\left\{ \nu ,c,f \right\}$ and the notation
\begin{equation}
s\left( {{\nu }_{1}},{{\nu }_{2}} \right)c\left( {{c}_{1}},{{c}_{2}} \right)a\left( {{f}_{1}},{{f}_{2}} \right)\equiv F\left( {{\xi }_{1}},{{\xi }_{2}} \right).
\label{eq:poznachennaF}	
\end{equation}
for the dependence of the probability amplitude on the internal indices. Then, instead of \eqref{eq:coordinatna_castina}, we may write
\begin{equation}
\begin{split}
\left| \mu  \right\rangle =F\left( {{\xi }_{1}},{{\xi }_{2}} \right)\int{d{{{\mathbf{r}}}_{2}}d{{{\mathbf{r}}}_{1}}\psi \left( \left| {{{\mathbf{r}}}_{2}}-{{{\mathbf{r}}}_{1}} \right| \right)}{{{\hat{q}}}^{+}}\left( {{\xi }_{1}},{{{\mathbf{r}}}_{1}} \right){{{\hat{\bar{q}}}}^{+}}\left( {{\xi }_{2}},{{{\mathbf{r}}}_{2}} \right)\left| 0 \right\rangle.
\end{split}
\label{eq:coord_cast}
\end{equation}
It is well known that, in the field theory, the operator ${{\hat{M}}_{03}}$ has the form: 
\begin{equation}
{{\hat{M}}_{03}}=\int{d\mathbf{r}}\left( {{x}_{3}}{{{\hat{T}}}_{00}}\left( {\mathbf{r}} \right)-{{x}_{0}}{{{\hat{T}}}_{30}}\left( {\mathbf{r}} \right) \right),
\label{eq:vt_kv_generator}	
\end{equation}
where ${{\hat{T}}_{00}}\left( {\mathbf{r}} \right)$ and ${{\hat{T}}_{30}}\left( {\mathbf{r}} \right)$ are the operators of corresponding component of the energy-momentum tensor, ${{x}_{0}}\equiv t$ is the time-like component of the coordinate 4-vector, and ${{x}_{3}}\equiv \left( -z \right)$ --  is its covariant component along covariant axis $OZ$. Therefore, the relation \eqref{eq:vt_kv_generator} can obviously be rewritten in the form
\begin{equation}
{{\hat{M}}_{03}}=-t{{\hat{P}}_{z}}+\int{d\mathbf{r}}\left( {{x}_{3}}{{{\hat{T}}}_{00}}\left( {\mathbf{r}} \right) \right),
\label{eq:M03_t}	
\end{equation}
where ${{\hat{P}}_{z}}$ is operator of the $z$ component of the total momentum of the system.

Note that relations \eqref{eq:vt_kv_generator} and \eqref{eq:M03_t} are exact and demand no assumptions and approximations. The dependence on $t$ in \eqref{eq:M03_t} coincides with dependence \eqref{eq:pripucenna1}. At the same time, the latter is a consequence of assumptions \eqref{eq:pripucenna} and \eqref{eq:pripucenna3}. Hence, we may draw conclusion that \eqref{eq:M03_t} proves the correctness of our assumptions.

State \eqref{eq:coord_cast}  is an eigenstate for the total momentum of the system and corresponds to its zero eigenvalue. Therefore, the action of the first term in \eqref{eq:M03_t} on this state is trivial and gives zero. So let us consider the second term in \eqref{eq:M03_t}. We introduce the following notation for it:
\begin{equation}
{{\hat{M}}_{03}}\left( {{{\hat{T}}}_{00}} \right)=\int{{{x}_{3}}}{{\hat{T}}_{00}}\left( {\mathbf{r}} \right)d\mathbf{r}.
\label{eq:M03_ot_T00}	
\end{equation}
In order to act by this operator on state of the two-particle system \eqref{eq:coord_cast}, we have to construct a non-relativistic approximation for the energy density ${{T}_{00}}\left( {\mathbf{r}} \right)$. The most convenient way to solve this problem is to use second-quantization representation, because Hamiltonian is represented here as an integral of a certain operator-valued function, which can be adopted as the non-relativistic limit of the energy density.

In the second-quantization representation, the non-relativistic Hamiltonian of a quark-antiquark system can be written in the form
\begin{equation}
\begin{split}
\hat{H}&={{{\hat{H}}}^{\left( 0 \right)}}+{{{\hat{H}}}^{\left( V \right)}}, \\
{{{\hat{H}}}^{\left( 0 \right)}}&=\int{d\mathbf{r}}\left( {{{\hat{\bar{q}}}}^{+}}\left( \xi ,\mathbf{r} \right)\left( -\frac{1}{2m}\Delta  \right){{{\hat{q}}}^{-}}\left( \xi ,\mathbf{r} \right) \right)\\
&+\int{d\mathbf{r}}\left( {{{\hat{q}}}^{+}}\left( \xi ,\vec{r} \right)\left( -\frac{1}{2m}\Delta  \right){{{\hat{\bar{q}}}}^{-}}\left( \xi ,\mathbf{r} \right) \right), \\
{{{\hat{H}}}^{\left( V \right)}}&=\frac{1}{2}\int{d{{{\mathbf{r}}}_{1}}d{{{\mathbf{r}}}_{2}}}V\left( {{{\mathbf{r}}}_{2}}-{{{\mathbf{r}}}_{1}} \right){{{\hat{\bar{q}}}}^{+}}\left( {{\xi }_{1}},{{{\mathbf{r}}}_{1}} \right)\\
&\times{{{\hat{q}}}^{+}}\left( {{\xi }_{2}},{{{\mathbf{r}}}_{2}} \right){{{\hat{\bar{q}}}}^{-}}\left( {{\xi }_{2}},{{{\mathbf{r}}}_{2}} \right){{{\hat{q}}}^{-}}\left( {{\xi }_{1}},{{{\mathbf{r}}}_{1}} \right).\\
\end{split}
\label{eq:H}
\end{equation}
Here, $V\left({{{\mathbf{r}}}_{2}}-{{{\mathbf{r}}}_{1}}\right)$ is a potential energy of interaction between the quark and antiquark, $m$ is the quark and antiquark mass. The latter parameter is approximately considered to be independent of the flavor, because the bound state exists owing to the strong interaction, whereas other types of interaction are neglected.

It is well-known that if we want to express the two-particle Hamiltonian in terms of differential operators, we can use the Jacobi coordinates,
\begin{equation}
\mathbf{R}=\frac{1}{2}\left( {{{\mathbf{r}}}_{1}}+{{{\mathbf{r}}}_{2}} \right),~~~\mathbf{r}={{\mathbf{r}}_{2}}-{{\mathbf{r}}_{1}},
\label{eq:Jacobi}	
\end{equation}
and present it as a sum of two commutative operators: the Hamiltonian of the center-of-mass, which depends only on $\mathbf{R}$; and the internal Hamiltonian, which depends only on $\mathbf{r}$. We want to obtain a similar form in the case where the Hamiltonian is written in terms of creation and annihilation operators. For this purpose, it is convenient to rewrite the one-particle part of the Hamiltonian $\hat{H}^{(0)}$ in the form of a twoparticle operator. Let us take into account that, in
the non-relativistic approximation, all operators can be considered in a subspace of the Fock space, where the number of particles and their content are fixed. In our case, we consider the subspace of states that contains one quark and one antiquark. The basis states in this subspace can be written in the form
\begin{equation}
\left| {{\xi }_{1}},{{\xi }_{2}},{{{\mathbf{r}}}_{1}},{{{\mathbf{r}}}_{2}} \right\rangle ={{\hat{\bar{q}}}^{+}}\left( {{\xi }_{1}},{{{\mathbf{r}}}_{1}} \right){{\hat{q}}^{+}}\left( {{\xi }_{2}},{{{\mathbf{r}}}_{2}} \right)\left| 0 \right\rangle.
\label{eq:bazis}	
\end{equation}
Acting by the operator
\begin{equation}
\hat{E}=\int{d\mathbf{r}}{{\hat{\bar{q}}}^{+}}\left( \xi ,\mathbf{r} \right){{\hat{q}}^{-}}\left( \xi ,\mathbf{r} \right),
\label{eq:edinichnij}	
\end{equation}
on any linear combination of states \eqref{eq:bazis}, one can get convinced that this operator plays the role of the identity operator in this subspace. The operator
\begin{equation}
{\hat{E}}'=\int{d\mathbf{r}}{{\hat{q}}^{+}}\left( \xi ,\mathbf{r} \right){{\hat{\bar{q}}}^{-}}\left( \xi ,\mathbf{r} \right).
\label{eq:edinichnij1}	
\end{equation}
has the same property in this subspace. If the first term in the one-particle part $\hat{H}^{(0)}$ of Hamiltonian \eqref{eq:H} is multiplied by the identity operator \eqref{eq:edinichnij1} and the second term by operator \eqref{eq:edinichnij}, we obtain an expression for the one-particle part in the apparently two-particle form,
\begin{equation}
\begin{split}
& {{{\hat{H}}}^{\left( 0 \right)}}=\left( -\frac{1}{2m} \right)\int{d{{{\mathbf{r}}}_{1}}d{{{\mathbf{r}}}_{2}}}\\
&\left( {{{\hat{\bar{q}}}}^{+}}\left( {{\xi }_{1}},{{{\mathbf{r}}}_{1}} \right){{{\hat{q}}}^{+}}\left( {{\xi }_{2}},{{{\mathbf{r}}}_{2}} \right){{\Delta }_{1}}{{{\hat{\bar{q}}}}^{-}}\left( {{\xi }_{2}},{{{\mathbf{r}}}_{2}} \right){{{\hat{q}}}^{-}}\left( {{\xi }_{1}},{{{\mathbf{r}}}_{1}} \right) \right. \\
& \left. +{{{\hat{\bar{q}}}}^{+}}\left( {{\xi }_{1}},{{{\mathbf{r}}}_{1}} \right){{{\hat{q}}}^{+}}\left( {{\xi }_{2}},{{{\mathbf{r}}}_{2}} \right){{\Delta }_{2}}{{{\hat{\bar{q}}}}^{-}}\left( {{\xi }_{2}},{{{\mathbf{r}}}_{2}} \right){{{\hat{q}}}^{-}}\left( {{\xi }_{1}},{{{\mathbf{r}}}_{1}} \right) \right).\\
\end{split}
\label{eq:odnodvux}	
\end{equation}
Now, let us replace the one-particle part of Hamiltonian \eqref{eq:H} by expression of \eqref{eq:odnodvux} and transform the result
to the Jacobi variables \eqref{eq:Jacobi}. Let us also introduce
the notations
\begin{equation}
\begin{split}
& {{{\mathbf{r}}}_{1}}\left( \mathbf{R},\mathbf{r} \right)=\mathbf{R}-\frac{1}{2}\mathbf{r},~~~{{{\mathbf{r}}}_{2}}\left( \mathbf{R},\mathbf{r} \right)=\mathbf{R}+\frac{1}{2}\mathbf{r}, \\
& {{{\hat{\bar{q}}}}^{+}}\left( {{\xi }_{1}},{{{\mathbf{r}}}_{1}}\left( \mathbf{R},\mathbf{r} \right) \right)={{{\hat{\bar{q}}}}^{+}_{1}},~~~{{{\hat{q}}}^{+}}\left( {{\xi }_{2}},{{{\mathbf{r}}}_{2}}\left( \mathbf{R},\mathbf{r} \right) \right)={{{\hat{q}}}^{+}_{2}}, \\
& {{{\hat{q}}}^{-}}\left( {{\xi }_{1}},{{{\mathbf{r}}}_{1}}\left( \mathbf{R},\mathbf{r} \right) \right)={{{\hat{q}}}^{-}}_{1}. 
~~~{{{\hat{\bar{q}}}}^{-}}\left( {{\xi }_{2}},{{{\mathbf{r}}}_{2}}\left( \mathbf{R},\mathbf{r} \right) \right)={{{\hat{\bar{q}}}}^{-}_{2}}.\\
\end{split}
\label{eq:pozn27_06}	
\end{equation}
Then instead of Hamiltonian \eqref{eq:H} we will get
\begin{equation}
\begin{split}
& \hat{H}={{{\hat{H}}}^{\left( {\mathbf{R}} \right)}}+{{{\hat{H}}}^{\left( \mathbf{r},V \right)}}, \\ 
& {{{\hat{H}}}^{\left( {\mathbf{R}} \right)}}=\left( -\frac{1}{4m} \right)\int{d\mathbf{R}d\mathbf{r}}{{{\hat{\bar{q}}}}^{+}_{1}}{{{\hat{q}}}^{+}_{2}}{{\Delta }_{{\mathbf{R}}}}{{{\hat{\bar{q}}}}^{-}_{2}}{{{\hat{q}}}^{-}_{1}}, \\ 
& {{{\hat{H}}}^{\left( \mathbf{r},V \right)}}=\int{d\mathbf{R}d\pmb{r}}{{{\hat{\bar{q}}}}^{+}_{1}}{{{\hat{q}}}^{+}}_{2}\left( -\frac{1}{m}{{\Delta }_{{\pmb{r}}}}+V\left( {\pmb{r}} \right) \right){{{\hat{\bar{q}}}}^{-}_{2}} {{{\hat{q}}}^{-}}_{1} .\\
\end{split}
\label{eq:Hvnut_vnech}
\end{equation}
The operator ${\hat{H}}^{\left( {\mathbf{R}} \right)}$ will be called the center-of-mass
Hamiltonian, and the operator ${{\hat{H}}^{\left( \mathbf{r},V \right)}}$ the internal
Hamiltonian of the system. Then the Hamiltonian $\hat{H}$ can be written in the form 
\begin{eqnarray}
\hat{H}=\int{{{{\hat{T}}}_{00}}\left( {\mathbf{R}} \right)d\mathbf{R}},
\label{eq:oprT00} 
\end{eqnarray}
where the energy density operator ${{\hat{T}}_{00}}\left( {\mathbf{R}} \right)$ can be written in the form:
\begin{equation}
{{\hat{T}}_{00}}\left( {\mathbf{R}} \right)=T_{00}^{\left( {\mathbf{R}} \right)}\left( {\mathbf{R}} \right)+T_{00}^{\left( {\mathbf{r}} \right)}\left( {\mathbf{R}} \right)+T_{00}^{\left( V \right)}\left( {\mathbf{R}} \right),
\label{eq:T00RrV}
\end{equation}
with the help of the following denotations:
\begin{equation}
\begin{split}
& T_{00}^{\left( {\mathbf{R}} \right)}\left( {\mathbf{R}} \right)=\left( -\frac{1}{4m} \right)\int{d\mathbf{r}}{{{\hat{\bar{q}}}}^{+}_{1}}{{{\hat{q}}}^{+}_{2}}{{\Delta }_{{\mathbf{R}}}}{{{\hat{\bar{q}}}}^{-}_{2}}{{{\hat{q}}}^{-}_{1}}, \\ 
& T_{00}^{\left( {\mathbf{r}} \right)}\left( {\mathbf{R}} \right)=\left( -\frac{1}{m} \right)\int{d\mathbf{r}}{{{\hat{\bar{q}}}}^{+}_{1}}{{{\hat{q}}}^{+}_{2}}{{\Delta }_{{\vec{r}}}}{{{\hat{\bar{q}}}}^{-}_{2}}{{{\hat{q}}}^{-}_{1}}, \\ 
& T_{00}^{\left( V \right)}\left( {\mathbf{R}} \right)=\int{d\mathbf{r}}{{{\hat{\bar{q}}}}^{+}_{1}}{{{\hat{q}}}^{+}_{2}}V\left( {\mathbf{r}} \right){{{\hat{\bar{q}}}}^{-}_{2}}{{{\hat{q}}}^{-}_{1}}.
\label{eq:T00RrVoboznacenija}
\end{split}
\end{equation}

The expressions in Eqs.(\ref{eq:oprT00})-(\ref{eq:T00RrVoboznacenija}) define the non-relativistic approximation of the operator ${{\hat{T}}_{00}}\left( {\mathbf{r}} \right)$ in second quantization  representation on Fock subspace.
It can be used to construct non-relativistic approximation for generator \eqref{eq:M03_ot_T00},
\begin{equation}
\begin{split}
&{{\hat{M}}_{03}}\left( {{{\hat{T}}}_{00}} \right)=\hat{M}_{03}^{\left( {\mathbf{R}} \right)}+\hat{M}_{03}^{\left( {\mathbf{r}} \right)}+\hat{M}_{03}^{\left( V \right)},\\
&\hat{M}_{03}^{\left( a \right)}=\int{d\mathbf{R}}\left( {{R}_{3}}T_{00}^{\left( a \right)}\left( {\mathbf{R}} \right) \right),
\label{eq:oprM03}
\end{split}
\end{equation}
where the index $a$ has three possible values: $a=\mathbf{R},\mathbf{r},V$.

Knowing the non-relativistic approximation for generator \eqref{eq:vt_kv_generator}, we may apply the associated operator \eqref{eq:Fokstan} to the non-relativistic approximation of state \eqref{eq:coord_cast} and obtain the probability amplitude for this state in the new reference frame. In order to simplify the action of the operator exponential function \eqref{eq:Fokstan} on \eqref{eq:coord_cast}, let us take into account that state \eqref{eq:coord_cast} is an eigenstate of the internal Hamiltonian ${{{\hat{H}}}^{\left( \mathbf{r},V \right)}}$. We are interested in the ground state of the system of bound quarks. In the corresponding center-of-mass reference frame, it corresponds to the eigenvalue equal to hadron’s mass. However, this eigenvalue is not degenerate for the ground state. Therefore, if we prove that generator \eqref{eq:vt_kv_generator} commutes with the internal Hamiltonian, this fact will mean that state \eqref{eq:coordinatna_castina} is also an eigenstate for both generator  \eqref{eq:vt_kv_generator} and operator \eqref{eq:Fokstan}.

Hence, we have to prove that the operators ${{\hat{M}}_{03}}\left( {{{\hat{T}}}_{00}} \right)$ and ${{\hat{H}}^{\left( \mathbf{r},V \right)}}$ commute with each other. For this purpose, it is convenient to change to the momentum representation and to differentiate the operator functions of coordinates that enter the Laplace operators in the expression for Hamiltonian \eqref{eq:Hvnut_vnech}. In order to change to the momentum representation, let us write operators \eqref{eq:Hvnut_vnech} and the potential energy as follows:
\begin{equation}
\begin{split}
& {{{\hat{\bar{q}}}}^{+}_{1}}=\frac{1}{{{\left( 2\pi  \right)}^{{3}/{2}\;}}}\int{d{{{\mathbf{p}}}_{1}}}{{{\hat{\bar{q}}}}^{+}}\left( {{\xi }_{1}},{{{\mathbf{p}}}_{1}} \right)\exp\left( i \left(\mathbf{p}_{1} \cdot \mathbf{r}_{1} \right) \right), \\ 
& {{{\hat{q}}}^{+}_{2}}=\frac{1}{{{\left( 2\pi  \right)}^{{3}/{2}\;}}}\int{d{{{\mathbf{p}}}_{2}}}{{{\hat{q}}}^{+}}\left( {{\xi }_{2}},{{{\mathbf{p}}}_{2}} \right)\exp\left( i \left(\mathbf{p}_{2} \cdot \mathbf{r}_{2} \right) \right), \\  
& {{{\hat{\bar{q}}}}^{-}_{2}}=\frac{1}{{{\left( 2\pi  \right)}^{{3}/{2}\;}}}\int{d{{{\mathbf{p}}}_{3}}}{{{\hat{\bar{q}}}}^{-}}\left( {{\xi }_{2}},{{{\mathbf{p}}}_{3}} \right)\exp\left( i \left(\mathbf{p}_{3} \cdot \mathbf{r}_{2} \right) \right), \\  
& {{{\hat{q}}}^{-}_{1}}=\frac{1}{{{\left( 2\pi  \right)}^{{3}/{2}\;}}}\int{d{{{\mathbf{p}}}_{4}}}{{{\hat{q}}}^{-}}\left( {{\xi }_{1}},{{{\mathbf{p}}}_{4}} \right)\exp\left( i \left(\mathbf{p}_{4} \cdot \mathbf{r}_{1} \right) \right), \\ 
\end{split} 	
\end{equation}
Here, ${{\hat{\bar{q}}}^{+}}\left( {{\xi }_{1}},{{{\mathbf{p}}}_{1}} \right),{{\hat{q}}^{+}}\left( {{\xi }_{2}},{{{\mathbf{p}}}_{2}} \right),{{\hat{\bar{q}}}^{-}}\left( {{\xi }_{2}},{{{\mathbf{p}}}_{3}} \right),{{\hat{q}}^{-}}\left( {{\xi }_{1}},{{{\mathbf{p}}}_{4}} \right)$ are the operators of quark creation and annihilation in momentum eigenstates.
Analogously, we can write the  potential energy
\begin{equation}
V\left( {\mathbf{r}} \right)=\frac{1}{{{\left( 2\pi  \right)}^{{3}/{2}\;}}}\int{d\mathbf{k}}V\left( {\mathbf{k}} \right)\exp \left( i\mathbf{k}\cdot\mathbf{r} \right).
\end{equation}
The notations $\mathbf{r}_{1}$ and $\mathbf{r}_{2}$ stand for the functions $\mathbf{r}_{1}(\mathbf{R},\mathbf{r})$ and $\mathbf{r}_{2}(\mathbf{R},\mathbf{r})$, which are determined by formula \eqref{eq:Hvnut_vnech}, but their arguments are omitted to make the expressions more compact. Substituting them into \eqref{eq:T00RrVoboznacenija} and carrying out the integration (for more details see Appendix~\ref{AppendixA_reply_part01}), we obtain the following expression for the internal Hamiltonian $\hat{H}(\mathbf{r})$:
\begin{equation}
\begin{split}
{{\hat{H}}^{\left( \mathbf{r},V \right)}}={{\hat{H}}^{\left( {\mathbf{r}} \right)}}+{{\hat{H}}^{\left( V \right)}}.
\end{split}
\end{equation}
where 
\begin{equation}
\begin{split}
{{{\hat{H}}}^{\left( {\mathbf{r}} \right)}}&=\int{T_{00}^{\left( {\mathbf{r}} \right)}\left( {\mathbf{R}} \right)d\mathbf{R}}=\int{d\mathbf{P}}d\mathbf{p}\left( \frac{{{{\mathbf{p}}}^{2}}}{m} \right)\\
&\times {{{\hat{\bar{q}}}}^{+}}\left( {{\xi }_{1}},{{{\mathbf{p}}}_{1}}=\frac{1}{2}\mathbf{P}-\mathbf{p} \right){{{\hat{q}}}^{+}}\left( {{\xi }_{2}},{{{\mathbf{p}}}_{2}}=\frac{1}{2}\mathbf{P}+\mathbf{p} \right)\\ 
&\times {{{\hat{\bar{q}}}}^{-}}\left( {{\xi }_{2}},{{{\mathbf{p}}}_{3}}=\frac{1}{2}\mathbf{P}+\mathbf{p} \right){{{\hat{q}}}^{-}}\left( {{\xi }_{2}},{{{\mathbf{p}}}_{4}}=\frac{1}{2}\mathbf{P}-\mathbf{p} \right), \\ 
{{{\hat{H}}}^{\left( V \right)}}&=\int{T_{00}^{\left( V \right)}\left( {\mathbf{R}} \right)d\mathbf{R}}=\frac{1}{{{\left( 2\pi  \right)}^{{3}/{2}\;}}}\int{d\mathbf{P}}d\mathbf{p}d\mathbf{k}\cdot V({\mathbf{k}})\\
& \times {{{\hat{\bar{q}}}}^{+}}\left( {{\xi }_{1}},{{{\mathbf{p}}}_{1}}=\frac{1}{2}\mathbf{P}-\mathbf{p} \right)  {{{\hat{q}}}^{+}}\left( {{\xi }_{2}},{{{\mathbf{p}}}_{2}}=\frac{1}{2}\mathbf{P}+\mathbf{p} \right)\\
&\times{{{\hat{\bar{q}}}}^{-}}\left( {{\xi }_{2}},{{{\mathbf{p}}}_{3}}=\frac{1}{2}\mathbf{P}+\mathbf{p}+\mathbf{k} \right){{{\hat{q}}}^{-}}\left( {{\xi }_{2}},{{{\mathbf{p}}}_{4}}=\frac{1}{2}\mathbf{P}-\mathbf{p}-\mathbf{k} \right). \\  
\end{split}
\end{equation}

Now, let us consider the expression for generator \eqref{eq:oprM03} in the momentum representation. For the term $\hat{M}_{03}^{\left( {\mathbf{R}} \right)}$, we have
\begin{equation}
\begin{split}
\hat{M}_{0,3}^{\left( {\mathbf{R}} \right)}&=\frac{1}{{{\left( 2\pi  \right)}^{3}}}\int{d{{{\mathbf{P}}}_{12}}d{{{\mathbf{P}}}_{34}}}d\mathbf{p}\left( \frac{{{\left( {{{\mathbf{P}}}_{34}} \right)}^{2}}}{4m} \right) \\ 
& \times{{{\bar{q}}}^{+}}\left( {{\xi }_{1}},{{{\mathbf{p}}}_{1}}=\frac{1}{2}{{{\mathbf{P}}}_{12}}-\mathbf{p} \right){{{\hat{q}}}^{+}}\left( {{\xi }_{2}},{{{\mathbf{p}}}_{2}}=\frac{1}{2}{{{\mathbf{P}}}_{12}}+\mathbf{p} \right) \\ 
& \times {{{\hat{\bar{q}}}}^{-}}\left( {{\xi }_{2}},{{{\mathbf{p}}}_{3}}=\frac{1}{2}{{{\mathbf{P}}}_{34}}+\mathbf{p} \right){{{\hat{q}}}^{-}}\left( {{\xi }_{2}},{{{\mathbf{p}}}_{4}}=\frac{1}{2}{{{\mathbf{P}}}_{34}}-\mathbf{p} \right)\\
&\times\int{{{R}_{3}}\exp \left( i\left( {{{\mathbf{P}}}_{12}}-{{{\mathbf{P}}}_{34}} \right)\mathbf{R} \right)d\mathbf{R}}. \\ 
\end{split}
\end{equation}
Here, the notations $\mathbf{p}_{1} + \mathbf{p}_{2} = \mathbf{P}_{12}$ and $\mathbf{p}_{3} +\mathbf{p}_{4} = \mathbf{P}_{34}$ for new integration variables were introduced.

The integration variable will be designated as ${\mathbf{P}}$ and, instead of $\mathbf{P}_{34}$, we introduce a new integration variable $\pmb{\varepsilon}$, by using the formula
\begin{equation}
{{\mathbf{P}}_{34}}=\mathbf{P}-\pmb{\varepsilon }.
\end{equation}
After required transformations, we obtain
\begin{equation}
\begin{split}
& \hat{M}_{03}^{\left( {\mathbf{R}} \right)}=-i\int{d\mathbf{P}d\pmb{\varepsilon }}d\mathbf{p}\left( \frac{\partial \delta \left( {\pmb{2\varepsilon }} \right)}{\partial {{\varepsilon }_{3}}} \right)\left( \frac{{{\left( \mathbf{P}-\pmb{\varepsilon } \right)}^{2}}}{4m} \right)\\
& \times {{{\bar{q}}}^{+}}\left( {{\xi }_{1}},{{{\mathbf{p}}}_{1}}=\frac{1}{2}\mathbf{P}-\mathbf{p} \right) {{{\hat{q}}}^{+}}\left( {{\xi }_{2}},{{{\mathbf{p}}}_{2}}=\frac{1}{2}\mathbf{P}+\mathbf{p} \right)\\
& \times {{{\hat{\bar{q}}}}^{-}}\left( {{\xi }_{2}},{{{\mathbf{p}}}_{3}}=\frac{1}{2}\left( \mathbf{P}-\pmb{\varepsilon } \right)+\mathbf{p} \right){{{\hat{q}}}^{-}}\left( {{\xi }_{2}},{{{\mathbf{p}}}_{4}}=\frac{1}{2}\left( \mathbf{P}-\pmb{\varepsilon } \right)-\mathbf{p} \right). 
\end{split}
\end{equation}
Here, the derivative of the Dirac $\delta$-function is understood in the sense usual for generalized functions, when integrating by parts Ref.\cite{Gelfand:1964}. Analogously,
\begin{equation}
\begin{split}
& \hat{M}_{03}^{\left( {\mathbf{r}} \right)}=-i\int{d\mathbf{P}d\pmb{\varepsilon }}d\mathbf{p}\left( \frac{\partial \delta \left( {\pmb{2\varepsilon }} \right)}{\partial {{\varepsilon }_{3}}} \right)\left( \frac{{{{\mathbf{p}}}^{2}}}{m} \right)\\
& \times {{{\bar{q}}}^{+}}\left( {{\xi }_{1}},{{{\mathbf{p}}}_{1}}=\frac{1}{2}\mathbf{P}-\mathbf{p} \right){{{\hat{q}}}^{+}}\left( {{\xi }_{2}},{{{\mathbf{p}}}_{2}}=\frac{1}{2}\mathbf{P}+\mathbf{p} \right) \\
& \times {{{\hat{\bar{q}}}}^{-}}\left( {{\xi }_{2}},{{{\mathbf{p}}}_{3}}=\frac{1}{2}\left(\mathbf{P}-\pmb{\varepsilon } \right)+\mathbf{p} \right) {{{\hat{q}}}^{-}}\left( {{\xi }_{2}},{{{\mathbf{p}}}_{4}}=\frac{1}{2}\left(\mathbf{P}-\pmb{\varepsilon } \right)-\mathbf{p} \right). \\ 
& \hat{M}_{03}^{\left( V \right)}=\frac{-i}{{{\left( 2\pi  \right)}^{{3}/{2}\;}}}\int{d\mathbf{P}d\pmb{\varepsilon }}d\mathbf{p}d\mathbf{k}\frac{\partial \delta \left( {\pmb{2\varepsilon }} \right)}{\partial {{\varepsilon }_{3}}}V\left( {\mathbf{k}} \right)\\
& \times {{{\hat{\bar{q}}}}^{+}}\left( {{\xi }_{1}},{{{\mathbf{p}}}_{1}}=\frac{1}{2}\mathbf{P}-\mathbf{p} \right){{{\hat{q}}}^{+}}\left( {{\xi }_{2}},{{{\mathbf{p}}}_{2}}=\frac{1}{2}\mathbf{P}+\mathbf{p} \right)\\
& \times {{{\hat{\bar{q}}}}^{-}}\left( {{\xi }_{2}},{{{\mathbf{p}}}_{3}}=\frac{1}{2}\left( \mathbf{P}-\pmb{\varepsilon } \right)+\mathbf{p}+\mathbf{k} \right)\\
&\times {{{\hat{q}}}^{-}}\left( {{\xi }_{2}},{{{\mathbf{p}}}_{4}}=\frac{1}{2}\left( \mathbf{P}-\pmb{\varepsilon } \right)-\mathbf{p}-\mathbf{k} \right).
\end{split}
\end{equation}

Knowing the expressions for the generator ${{\hat{M}}_{03}}$ and the internal Hamiltonian, we can calculate their commutator. Since each of those operators consists of several terms, let us consider commutators between those terms. While calculating the commutators, it is convenient firstly to convert the products of operators arranged in that or another order into the normal form with the help of Wick's theorem. 

Above, we have already used the fact that all operators are considered in a subspace of the Fock space, whose states include one quark and one antiquark. Therefore, the operators containing, in the normal form, two or more quark/antiquark creation or annihilation operators will have zero matrix elements for all basis elements of this subspace, so that such operators can be dropped. Taking into account that each of the operators concerned contains one quark and one antiquark operators, their product will include two quark and two antiquark operators, thus containing “redundant” operators. Applying Wick's theorem to this product, we obtain that only those terms will have nonzero matrix elements in the space concerned, in which the “redundant” operators are paired.

As an example, let us analyze the product $\hat{H}^{\left( {\mathbf{r}} \right)}\hat{M}_{03}^{\left(\mathbf{R} \right)}$. After converting it to the normal form, dropping the terms with zero matrix elements, and considering the $\delta$-functions that arise as a result of the pairing, this product can be written in the following form (more detailed calculations can be found in Appendix~\ref{AppendixA_reply_part01}:
\begin{equation}
\begin{split}
& {{{\hat{H}}}^{\left( {\mathbf{r}} \right)}}\hat{M}_{03}^{\left( {\mathbf{R}} \right)}=-i\int{d{{{\mathbf{p}}}_{1}}}d{{{\mathbf{p}}}_{2}}d\pmb{\varepsilon }\left( \frac{\partial \delta \left( {\pmb{2\varepsilon }} \right)}{\partial {{\varepsilon }_{3}}} \right)\\
&\times\left( \frac{{{\left( {{{\mathbf{p}}}_{2}}-{{{\mathbf{p}}}_{1}} \right)}^{2}}}{4m} \right)\left( \frac{{{\left( {{{\mathbf{p}}}_{1}}+{{{\mathbf{p}}}_{2}}-\pmb{\varepsilon } \right)}^{2}}}{4m} \right) \\ 
& \times {{{\hat{\bar{q}}}}^{+}}\left( {{\xi }_{1}},{{{\mathbf{p}}}_{1}} \right){{{\hat{q}}}^{+}}\left( {{\xi }_{2}},{{{\mathbf{p}}}_{2}} \right){{{\hat{\bar{q}}}}^{-}}\left( {{\xi }_{2}},{{{\mathbf{p}}}_{2}}-\frac{1}{2}\pmb{\varepsilon } \right){{{\hat{q}}}^{-}}\left( {{\xi }_{1}},{{{\mathbf{p}}}_{1}}-\frac{1}{2}\pmb{\varepsilon } \right). \\ 
\end{split}
\label{eq:Rrupr}
\end{equation} 
For the product of the same operators, but in the inverse order, after similar transformations, we obtain
\begin{equation}
\begin{split}
\hat{M}_{03}^{\left( {\mathbf{R}} \right)}{{{\hat{H}}}^{\left( {\mathbf{r}} \right)}}&=-i\int{d{{{\mathbf{p}}}_{1}}d{{{\mathbf{p}}}_{2}}}d{{{\mathbf{p}}}_{3}}d{{{\mathbf{p}}}_{4}}d\pmb{\varepsilon }\\
&\times \left( \frac{{{\left( {{{\mathbf{p}}}_{3}}-{{{\mathbf{p}}}_{4}} \right)}^{2}}}{4m} \right)\left( \frac{{{\left( {{{\mathbf{p}}}_{1}}+{{{\mathbf{p}}}_{2}}-\pmb{\varepsilon } \right)}^{2}}}{4m} \right)\left( \frac{\partial \delta \left( {\pmb{2\varepsilon }} \right)}{\partial {{\varepsilon }_{3}}} \right) \\ 
& \times \delta \left( \left( {{{\vec{p}}}_{2}}-\frac{1}{2}\pmb{\varepsilon } \right)-{{{\mathbf{p}}}_{3}} \right)\delta \left( \left( {{{\mathbf{p}}}_{1}}-\frac{1}{2}\pmb{\varepsilon } \right)-{{{\mathbf{p}}}_{4}} \right)\\
&\times{{{\bar{q}}}^{+}}\left( {{\xi }_{1}},{{{\mathbf{p}}}_{1}} \right){{{\hat{q}}}^{+}}\left( {{\xi }_{2}},{{{\mathbf{p}}}_{2}} \right){{{\hat{\bar{q}}}}^{-}}\left( {{\xi }_{2}},{{{\mathbf{p}}}_{3}} \right){{{\hat{q}}}^{-}}\left( {{\xi }_{1}},{{{\mathbf{p}}}_{4}} \right). \\  
\end{split}
\end{equation}
Carrying out the integration over the components ${{\mathbf{p}}_{3}}$ and ${{\mathbf{p}}_{4}}$, we obtain a result that is identical to expression \eqref{eq:Rrupr}.
Analogously, it can be demonstrated that the operator $\hat{M}_{03}^{\left( {\mathbf{R}} \right)}$ commutes with other terms in the internal Hamiltonian (for details see Appendix~\ref{AppendixA_reply_part01}). As a result, it commutes with the entire internal Hamiltonian of the bound quark system.

Now, let us calculate the commutators of the operator $\hat{M}_{03}^{\left( {\mathbf{r}} \right)}$ with the terms in the internal Hamiltonian. With the help of transformations similar to those considered above, the product ${{{\hat{H}}}^{\left( {\mathbf{r}} \right)}}\hat{M}_{03}^{\left( {\mathbf{r}} \right)}$ is obtained in the form
\begin{equation}
\begin{split}
{{{\hat{H}}}^{\left( {\mathbf{r}} \right)}}\hat{M}_{03}^{\left( {\mathbf{r}} \right)}&=-i\int{d{{{\mathbf{p}}}_{1}}d{{{\mathbf{p}}}_{2}}d{{{\mathbf{p}}}_{3}}d{{{\mathbf{p}}}_{4}}d\pmb{\varepsilon }} \\
&\times \left( \frac{\partial \delta \left( {\pmb{2\varepsilon }} \right)}{\partial {{\varepsilon }_{3}}} \right)\left( \frac{{{\left( {{{\mathbf{p}}}_{2}}-{{{\mathbf{p}}}_{1}} \right)}^{2}}}{4m} \right)\left( \frac{{{\left( {{{\mathbf{p}}}_{2}}-{{{\mathbf{p}}}_{1}} \right)}^{2}}}{4m} \right) \\ 
&\times {{{\hat{\bar{q}}}}^{+}}\left( {{\xi }_{1}},{{{\mathbf{p}}}_{1}} \right){{{\hat{q}}}^{+}}\left( {{\xi }_{2}},{{{\mathbf{p}}}_{2}} \right)\\
&\times {{{\hat{\bar{q}}}}^{-}}\left( {{\xi }_{2}},{{{\mathbf{p}}}_{3}}={{{\mathbf{p}}}_{2}}-\frac{1}{2}\pmb{\varepsilon } \right)\\
&\times {{{\hat{q}}}^{-}}\left( {{\xi }_{1}},{{{\mathbf{p}}}_{4}}={{{\mathbf{p}}}_{1}}-\frac{1}{2}\pmb{\varepsilon } \right). \\  
\end{split}
\end{equation}
The same result is obtained, if the product of the same operators, but taken in the inverse order, is converted to the normal form.  Thus, $\hat{M}_{03}^{\left( \mathbf{r} \right)}$ and $\hat{H}^{\left( {\mathbf{r}} \right)}$ commute with each other.

The calculation of the product ${{\hat{H}}^{\left( V \right)}}\hat{M}_{03}^{\left( {\mathbf{r}} \right)}$ leads to result
\begin{equation}
\begin{split}
{{{\hat{H}}}^{\left( V \right)}}\hat{M}_{03}^{\left( {\mathbf{r}} \right)}&=\frac{-i}{{{\left( 2\pi  \right)}^{{3}/{2}\;}}}\int{d{{{\mathbf{p}}}_{1}}d{{{\mathbf{p}}}_{2}}d\pmb{\varepsilon }}d\mathbf{k}\\
&\times \left( \frac{\partial \delta \left( {\pmb{2\varepsilon }} \right)}{\partial {{\varepsilon }_{3}}} \right) V\left( {\mathbf{k}} \right)\left( \frac{{{\left( {{{\mathbf{p}}}_{2}}-{{{\mathbf{p}}}_{1}}+2\mathbf{k} \right)}^{2}}}{4m} \right) \\ 
& \times {{{\hat{\bar{q}}}}^{+}}\left( {{\xi }_{1}},{{{\mathbf{p}}}_{1}} \right){{{\hat{q}}}^{+}}\left( {{\xi }_{2}},{{{\mathbf{p}}}_{2}} \right)\\
&\times {{{\hat{\bar{q}}}}^{-}}\left( {{\xi }_{2}},{{{\mathbf{p}}}_{3}}={{{\mathbf{p}}}_{2}}+\mathbf{k}-\frac{1}{2}\pmb{\varepsilon } \right)\\
&\times{{{\hat{q}}}^{-}}\left( {{\xi }_{1}},{{{\mathbf{p}}}_{4}}={{{\mathbf{p}}}_{1}}-\mathbf{k}-\frac{1}{2}\pmb{\varepsilon } \right). \\ 
\end{split}
\label{eq:HVMrpram}
\end{equation}
However, the calculation of the product $\hat{M}_{03}^{\left( {\mathbf{r}} \right)}{{\hat{H}}^{\left( V \right)}}$ gives a result that does not coincide with formula \eqref{eq:HVMrpram}:
\begin{equation}
\begin{split}
\hat{M}_{03}^{\left( {\mathbf{r}} \right)}{{{\hat{H}}}^{\left( V \right)}}&=\frac{-i}{{{\left( 2\pi  \right)}^{{3}/{2}\;}}}\int{d{{{\mathbf{p}}}_{1}}d{{{\mathbf{p}}}_{2}}d\pmb{\varepsilon }}d\mathbf{k}\\
&\times V\left( {\mathbf{k}} \right)\left( \frac{\partial \delta \left( {\pmb{2\varepsilon }} \right)}{\partial {{\varepsilon }_{3}}} \right)\left( \frac{{{\left( {{{\mathbf{p}}}_{2}}-{{{\mathbf{p}}}_{1}} \right)}^{2}}}{4m} \right) \\ 
&\times {{{\bar{q}}}^{+}}\left( {{\xi }_{1}},{{{\mathbf{p}}}_{1}} \right) {{{\hat{q}}}^{+}}\left( {{\xi }_{2}},{{{\mathbf{p}}}_{2}} \right)\\
&\times {{{\hat{\bar{q}}}}^{-}}\left( {{\xi }_{2}},{{{\mathbf{p}}}_{3}}={{{\mathbf{p}}}_{2}}+\mathbf{k}-\frac{1}{2}\pmb{\varepsilon } \right)\\
&\times {{{\hat{q}}}^{-}}\left( {{\xi }_{2}},{{{\mathbf{p}}}_{4}}={{{\mathbf{p}}}_{1}}-\mathbf{k}-\frac{1}{2}\pmb{\varepsilon } \right). \\ 
\end{split}
\label{eq:HVMrobr}
\end{equation}
Hence, the operators $\hat{M}_{03}^{\left( {\mathbf{r}} \right)}$ and ${{\hat{H}}^{\left( V \right)}}$ do not commute with each other. At the same time,
\begin{equation}
\begin{split}
\hat{M}_{03}^{\left( V \right)}{{{\hat{H}}}^{\left( {\mathbf{r}} \right)}}&=\frac{-i}{{{\left( 2\pi  \right)}^{{3}/{2}\;}}}\int{d{{{\mathbf{p}}}_{1}}d{{{\mathbf{p}}}_{2}}d{{{\mathbf{p}}}_{3}}d{{{\mathbf{p}}}_{4}}d\pmb{\varepsilon }}d\mathbf{k}\\
&\times \left( \frac{\partial \delta \left( {\pmb{2\varepsilon }} \right)}{\partial {{\varepsilon }_{3}}} \right) V\left( {\mathbf{k}} \right)\left( \frac{{{\left( {{{\mathbf{p}}}_{2}}-{{{\mathbf{p}}}_{1}}+2\mathbf{k} \right)}^{2}}}{4m} \right) \\ 
& \times {{{\hat{\bar{q}}}}^{+}}\left( {{\xi }_{1}},{{{\mathbf{p}}}_{1}} \right){{{\hat{q}}}^{+}}\left( {{\xi }_{2}},{{{\mathbf{p}}}_{2}} \right)\\
&\times {{{\hat{\bar{q}}}}^{-}}\left( {{\xi }_{2}},{{{\mathbf{p}}}_{3}}={{{\mathbf{p}}}_{2}}+\mathbf{k}-\frac{1}{2}\pmb{\varepsilon } \right)\\
&\times {{{\hat{q}}}^{-}}\left( {{\xi }_{1}},{{{\mathbf{p}}}_{4}}={{{\mathbf{p}}}_{1}}-\mathbf{k}-\frac{1}{2}\pmb{\varepsilon } \right). \\ 
\end{split}
\label{eq:MVHrpram}
\end{equation}
and 
\begin{equation}
\begin{split}
{{{\hat{H}}}^{\left( {\mathbf{r}} \right)}}\hat{M}_{03}^{\left( V \right)}&=\frac{-i}{{{\left( 2\pi  \right)}^{{3}/{2}\;}}}\int{d{{{\mathbf{p}}}_{1}}d{{{\mathbf{p}}}_{2}}d{{{\mathbf{p}}}_{3}}d{{{\mathbf{p}}}_{4}}}d\pmb{\varepsilon }d\mathbf{k}\\
&\times \left( \frac{\partial \delta \left( {\pmb{2\varepsilon }} \right)}{\partial {{\varepsilon }_{3}}} \right) V\left( {\mathbf{k}} \right)\left( \frac{{{\left( {{{\mathbf{p}}}_{2}}-{{{\mathbf{p}}}_{1}} \right)}^{2}}}{m} \right) \\ 
& \times {{{\hat{\bar{q}}}}^{+}}\left( {{\xi }_{1}},{{{\mathbf{p}}}_{1}} \right){{{\hat{q}}}^{+}}\left( {{\xi }_{2}},{{{\mathbf{p}}}_{2}} \right)\\
&\times {{{\hat{\bar{q}}}}^{-}}\left( {{\xi }_{2}},{{{\mathbf{p}}}_{3}}={{{\mathbf{p}}}_{2}}+\mathbf{k}-\frac{1}{2}\pmb{\varepsilon } \right)\\
&\times {{{\hat{q}}}^{-}}\left( {{\xi }_{1}},{{{\mathbf{p}}}_{4}}={{{\mathbf{p}}}_{1}}-\mathbf{k}-\frac{1}{2}\pmb{\varepsilon } \right). \\ 
\end{split}
\label{eq:MVHobr}
\end{equation}

Taking into account that the expressions for ${{\hat{H}}^{\left( V \right)}}\hat{M}_{03}^{\left( {\mathbf{r}} \right)}$ and $\hat{M}_{03}^{\left( V \right)}{{\hat{H}}^{\left( {\mathbf{r}} \right)}}$ [Eqs.(\ref{eq:HVMrpram}) and (\ref{eq:MVHrpram}), respectively] enter the general expression for the commutator $\left[ {{{\hat{H}}}^{\left( \mathbf{r},V \right)}},{{{\hat{M}}}_{03}} \right]$ with opposite signs (the same can be said about Eqs.(\ref{eq:HVMrobr}) and (\ref{eq:MVHobr})), we arrive at a conclusion that
\begin{equation}
\left[ {{{\hat{H}}}^{\left( {\mathbf{r}} \right)}},\hat{M}_{03}^{\left( V \right)} \right]+\left[ {{{\hat{H}}}^{\left( V \right)}},\hat{M}_{03}^{\left( {\mathbf{r}} \right)} \right]=0.
\end{equation}
Finally, computation of products $\hat{M}_{03}^{\left( V \right)}{{\hat{H}}^{\left( V \right)}} $ and ${{\hat{H}}^{\left( V \right)}}\hat{M}_{03}^{\left( V \right)}$ give us the same result:
\begin{equation}
\begin{split}
\hat{M}_{03}^{\left( V \right)}{{{\hat{H}}}^{\left( V \right)}}&={{{\hat{H}}}^{\left( V \right)}}\hat{M}_{03}^{\left( V \right)}\\
&=\frac{-i}{{{\left( 2\pi  \right)}^{3}}}\int{d{{{\mathbf{p}}}_{1}}d{{{\mathbf{p}}}_{2}}}d\pmb{\varepsilon }d{\mathbf{p}}'d{\mathbf{k}}'\frac{\partial \delta \left( {\pmb{2\varepsilon }} \right)}{\partial {{\varepsilon }_{3}}}V\left( {\mathbf{k}} \right)V\left( {{\mathbf{k}}'} \right) \\ 
& \times {{{\hat{\bar{q}}}}^{+}}\left( {{\xi }_{1}},{{{\mathbf{p}}}_{1}} \right){{{\hat{q}}}^{+}}\left( {{\xi }_{2}},{{{\mathbf{p}}}_{2}} \right)\\
&\times {{{\hat{\bar{q}}}}^{-}}\left( {{\xi }_{2}},{{{\mathbf{p}}}_{3}}={{{\mathbf{p}}}_{2}}+{\mathbf{k}}'+\mathbf{k}-\frac{1}{2}\pmb{\varepsilon } \right)\\
&\times {{{\hat{q}}}^{-}}\left( {{\xi }_{1}},{{{\mathbf{p}}}_{4}}={{{\mathbf{p}}}_{1}}-{\mathbf{k}}'-\mathbf{k}-\frac{1}{2}\pmb{\varepsilon } \right). \\ 
\end{split}
\end{equation}
Therefore, if we decompose the commutator $\left[ {{{\hat{H}}}^{\left( \mathbf{r},V \right)}},{{{\hat{M}}}_{03}} \right]$  into the terms that correspond to the terms of the internal Hamiltonian and the generator, the sum of all those terms will be equal to zero, i.e.
\begin{equation}
\left[ {{{\hat{M}}}_{03}},{{{\hat{H}}}^{\left( \mathbf{r},V \right)}} \right]=0.
\end{equation}

Thus, as was marked above, since state \eqref{eq:coord_cast} is an eigenstate of the internal Hamiltonian and corresponds to a nondegenerate eigenvalue, the consequence of relation \eqref{eq:vt_kv_generator} consists in that this state has to be an eigenstate of the boost generator as well:
\begin{equation}
{{\hat{M}}_{03}}\left| \mu  \right\rangle ={{m}_{03}}\left| \mu  \right\rangle,
\label{eq:eigenval}
\end{equation}
where ${{m}_{03}}$ is the eigenvalue corresponding to ${{\hat{M}}_{03}}$ - generator's eigenstate $\left| \mu  \right\rangle$. In order to determine this parameter, let us take advantage of the symmetry properties of the eigenstate $\left| \mu  \right\rangle$. In particular, this state must transform into itself at any inversion of coordinate axes. Furthermore, if the interaction potential between the quark and the antiquark is assumed to be spherically symmetric, the ground state of this system has also to be spherically symmetric, i.e. to transform into itself at any rotation. Using the notation ${{\hat{U}}^{\left( I,R \right)}}$ for the unitary operator of inversion or rotation in the considered subspace of the Fock space, we may write
\begin{equation}
{{\hat{U}}^{\left( I,R \right)}}\left| \mu  \right\rangle =\left| \mu  \right\rangle .
\label{eq:UIR}
\end{equation}
Then \eqref{eq:eigenval} can be rewritten in the form
\begin{equation}
{{\hat{M}}_{03}}{{\hat{U}}^{\left( I,R \right)}}\left| \mu  \right\rangle ={{m}_{03}}{{\hat{U}}^{\left( I,R \right)}}\left| \mu  \right\rangle,
\label{eq:eigenval1}
\end{equation}
or 
\begin{equation}
{{\left( {{{\hat{U}}}^{\left( I,R \right)}} \right)}^{-1}}{{\hat{M}}_{03}}{{\hat{U}}^{\left( I,R \right)}}\left| \mu  \right\rangle ={{m}_{03}}\left| \mu  \right\rangle.
\label{eq:eigenval-1}
\end{equation} 
The operator $\left( \hat{U}^{\left( I,R \right)} \right)^{-1}{{\hat{M}}_{03}}{{\hat{U}}^{\left( I,R \right)}}$ associated with ${\hat{M}}_{03}$ by means of the tensor transformation rule. This fact means, by selecting an inversion or a rotation that changes the $OZ$-axis direction to the opposite one, we obtain 
\begin{equation}
{{\left( {{{\hat{U}}}^{\left( I,R \right)}} \right)}^{-1}}{{\hat{M}}_{03}}{{\hat{U}}^{\left( I,R \right)}}=-{{\hat{M}}_{03}}.
\label{eq:minusM03}
\end{equation}

On the other hand, substituting \eqref{eq:minusM03} into \eqref{eq:eigenval-1} and taking \eqref{eq:eigenval} into account, we obtain
\begin{equation}
{{m}_{03}}=0.
\label{eq:ravnonulu}
\end{equation}

Hence, if $\left| {{\mu }'} \right\rangle $ stands for the state of a system of two bound particles in the reference frame that is obtained from the c.m.s. reference frame for those particles by means of a boost with rapidity $Y$ along the axis $OZ$, we have
\begin{equation}
\left| {{\mu }'} \right\rangle =\exp \left( i{{{\hat{M}}}_{03}}Y \right)\left| \mu  \right\rangle. 
\label{eq:mustrih}
\end{equation}
However, with regard for \eqref{eq:eigenval} and \eqref{eq:ravnonulu}, one can see that, of all
of all terms in the series representing the operator exponential function $\exp \left( i{{{\hat{M}}}_{03}}Y \right)$, only the term with the identity operator provides a non-zero result after its action on the state $\left|\mu\right\rangle$. Therefore, we obtain
\begin{equation}
\left| {{\mu }'} \right\rangle =\left| \mu  \right\rangle. 
\label{eq:mustrihravnomu}
\end{equation}
This result coincides with result \eqref{eq:visnovok} obtained above in terms of differential operators. Hence, a conclusion can be drawn that the internal state of a non-relativistic system consisting of bound particles does not vary in the case of a boost-like change to the reference frame, in which this bound system has a relativistic energy-momentum. Note that this conclusion has already been made in the literature Ref.\cite{Karmanov:1988}, but without any substantiation.

\section {The Group-Theory Analysis of the State Transformation Problem}

In the previous sections, we have examined two different representations of the boost generator ${{\hat{M}}_{03}}$ and obtained similar results. Therefore, a question arises: Whether these results can be generalized? The generalization can be reached, if we analyze the problem in the framework of the general group theory.

Let us consider the generators of Poincare group. There are four generators of space-time translations ${{\hat{P}}_{a}}(a=0,1,2,3)$ and six Lorentz generators ${{\hat{M}}_{ab}}=-{{\hat{M}}_{ba}}$. The commutation relations between these generators depend only on the group multiplication rule, so that those features of a state transformation in quantum-mechanical systems of interacting particles that were discussed above do not affect them. Furthermore, those commutation relations are independent of the generator representation. Therefore, they can be considered, knowing nothing about the explicit generator forms. It is known from Ref.\cite{Raider} that the operator ${{g}^{ab}}{{\hat{P}}_{a}}{{\hat{P}}_{b}}$ commute with every generator of the Poincare group and, in particular, with the generator ${{\hat{M}}_{03}}$, which is of interest for us. Taking into account that in accordance with the field quantization postulate Ref.\cite{Bogolubov:1959}, the generator ${{\hat{P}}_{0}}$ must coincide with the total Hamiltonian of the system, and the operators ${{\hat{P}}_{b}}(b=0,1,2,3)$ with the operators of momentum components. Then one may note that operator ${{g}^{ab}}{{\hat{P}}_{a}}{{\hat{P}}_{b}}$ is identical to the squared internal Hamiltonian of the system, since all of its eigenvalues are equal to the squared eigenvalues of the internal energy for the particle system under considation.

Hence, the boost generator commutes with the square of the internal Hamiltonian, irrespective of both the representation of those operators and a capability to apply the non-relativistic approximation in the center-of-mass reference frame of the particles concerned. If this approximation is applicable--it is so in the case that we are interested in--the eigenstate associated with the smallest eigenvalue of the squared internal Hamiltonian corresponds to a nondegenerate eigenvalue. Then, owing to the commutativity of the operators ${{g}^{ab}}{{\hat{P}}_{a}}{{\hat{P}}_{b}}$ and ${{\hat{M}}_{03}}$, we obtain that this state is an eigenstate for ${{\hat{M}}_{03}}$ as well. From whence, it follows that this state is not transformed at the boost. Hence, it becomes evident that the nondegenerate character of the ground state of a bound system plays the most important role.

Therefore, from the commutation relations between the generators of the Poincare group, it follows that, if the internal state of a system of interacting particles (i.e. the eigenstate of the squared internal Hamiltonian of the system) is not degenerate, it will not change at the boost transformation.

\section{Discussion of the Results and Conclusions}

The ground state of a system with central interaction is known to be spherically symmetric. According to the results obtained above, it will not transform, if we switch to another reference frame. In other words, it will remain spherically symmetric and will not undergo the Lorentz contraction. As was marked above, a conclusion similar to ours has been made rather long ago in the literature Ref.\cite{Karmanov:1988}. But the cited work was devoted to other issues, and the substantiation of this statement was not the aim of its author.

We understand that our conclusion about the absence of contraction contradicts the standard approach.In the majority of works (see, e.g., Refs.\cite{XiandongJi:2004, RevModPhys.84.1231, Dremin:2013}), their authors use a scenario that hadron's shape varies when changing from one inertial reference frame to another one. The same approach is also used in geometrical models (see, e.g., Ref.\cite{Conceicao201258}), which are based on the assumption that a hadron can be imagined as a “black disk” after the Lorentz contraction.

Concerning this contradiction, we would like to make the following remarks. First, our result does not contradict the relativity theory. To illustrate this, in \cite{Sharph:2014nza}), we considered a problem of the dependence of the distance between two classical (not quantummechanical) relativistic particles that move, being driven by the laws defined in their center-of-mass reference frame. The time dependence of the distance between the same particles, but in another inertial reference frame, can obviously be calculated with the help of only Lorentz transformations. In 
Appendix~\ref{AppendixA_reply_part01}, it was shown that we may assign such a law of particle motion that the observers in different inertial reference frames will measure the same time dependence of the distance in the corresponding “own” reference frames. So, the contraction does not occur in this problem, being also a consequence of only Lorentz transformations, as well as the contraction of rod’s length.

Second, the internal state of the rod was never examined in the problem about its motion. Of course, this state has nothing in common with the states considered in this work. Therefore, the results of this work are not related to the transformation of rod’s state. The problem about such states and their transformation goes far beyond the scope of issues that were covered in this work.

Third, one more argument in favor of our conclusion, in addition to the detailed arguments given above, can be put forward “on the basis of general considerations”. If we consider a meson with the zero spin, its internal angular momentum must equal zero. As a result, this state must be spherically symmetric. Furthermore, the internal angular momentum of the meson must remain equal to zero in any reference frame. However, if meson’s state had been no more spherically symmetric due to the Lorentz contraction, the internal angular momentum would have ceased to equal zero. In other words, the meson would have possessed a nonzero spin in the new reference frame.

The result obtained in this work is important for the further consideration. In particular, it allows the method of many-particle fields applied to the description of hadrons in scattering processes Refs.\cite{Volkotrub:2016, Chudak:2016} to be developed further. This method makes it possible to consider the scattering processes of hadrons as many quark systems and to describe the confinement of quarks and gluons.

\appendix
\numberwithin{equation}{section}
\numberwithin{figure}{section}
\numberwithin{table}{section}
\section{Appendix: The Commutation Relation Between $\hat{M}_{03}(\hat{T}_{00})$ and $\hat{H}^{({\mathbf{r}},V)}$}
\label{AppendixA_reply_part01}

In this section we would like to prove that the operators ${{\hat{M}}_{03}}\left( {{{\hat{T}}}_{00}} \right)$ and ${{\hat{H}}^{\left( \mathbf{r},V \right)}}$ commute, where the internal Hamiltonian is represented by the following expression:
\begin{equation}
\begin{split}
{{\hat{H}}^{\left( \mathbf{r},V \right)}}={{\hat{H}}^{\left( {\mathbf{r}} \right)}}+{{\hat{H}}^{\left( V \right)}}.
\end{split}
\label{eq:vnutrH1}
\end{equation}

To prove it, it is convenient to use the momentum representation for differentiation operator functions of the coordinates, which are included in the Laplace operator in the expression for the Hamiltonian \eqref{eq:Hvnut_vnech}. The operators \eqref{eq:pozn27_06} for transition to the momentum representation we will write in the form:
\begin{equation}
\begin{split}
& {{{\hat{\bar{q}}}}^{+}_{1}}=\frac{1}{{{\left( 2\pi  \right)}^{{3}/{2}\;}}}\int{d{{{\mathbf{p}}}_{1}}}{{{\hat{\bar{q}}}}^{+}}\left( {{\xi }_{1}},{{{\mathbf{p}}}_{1}} \right)\exp \left( i{{{\mathbf{p}}}_{1}}\left( \mathbf{R}-\frac{1}{2}\mathbf{r} \right) \right), \\ 
& {{{\hat{q}}}^{+}_{2}}=\frac{1}{{{\left( 2\pi  \right)}^{{3}/{2}\;}}}\int{d{{{\mathbf{p}}}_{2}}}{{{\hat{q}}}^{+}}\left( {{\xi }_{2}},{{{\mathbf{p}}}_{2}} \right)\exp \left( i{{{\mathbf{p}}}_{2}}\left( \mathbf{R}+\frac{1}{2}\mathbf{r} \right) \right), \\ 
& {{{\hat{\bar{q}}}}^{-}_{2}}=\frac{1}{{{\left( 2\pi  \right)}^{{3}/{2}\;}}}\int{d{{{\mathbf{p}}}_{3}}}{{{\hat{\bar{q}}}}^{-}}\left( {{\xi }_{2}},{{{\mathbf{p}}}_{3}} \right)\exp \left( -i{{{\mathbf{p}}}_{3}}\left( \mathbf{R}+\frac{1}{2}\mathbf{r} \right) \right), \\ 
& {{{\hat{q}}}^{-}_{1}}=\frac{1}{{{\left( 2\pi  \right)}^{{3}/{2}\;}}}\int{d{{{\mathbf{p}}}_{4}}}{{{\hat{q}}}^{-}}\left( {{\xi }_{1}},{{{\mathbf{p}}}_{4}} \right)\exp \left( -i{{{\mathbf{p}}}_{4}}\left( \mathbf{R}-\frac{1}{2}\mathbf{r} \right) \right), \\
\end{split} 	
\label{eq:impulspred}
\end{equation}
where ${{\hat{\bar{q}}}^{+}}\left( {{\xi }_{1}},{{{\mathbf{p}}}_{1}} \right),{{\hat{q}}^{+}}\left( {{\xi }_{2}},{{{\mathbf{p}}}_{2}} \right),{{\hat{\bar{q}}}^{-}}\left( {{\xi }_{2}},{{{\mathbf{p}}}_{3}} \right),{{\hat{q}}^{-}}\left( {{\xi }_{1}},{{{\mathbf{p}}}_{4}} \right)$ are the operators of quark creation and annihilation in momentum eigenstates. Analogously, we can write the  potential energy
\begin{equation}
V\left( {\mathbf{r}} \right)=\frac{1}{{{\left( 2\pi  \right)}^{{3}/{2}\;}}}\int{d\mathbf{k}}V\left( {\mathbf{k}} \right)\exp \left( i\mathbf{k}\mathbf{r} \right).
\label{eq:potential}
\end{equation}

Substituting them into \eqref{eq:T00RrVoboznacenija} we will get the integrals over the variables ${{\mathbf{p}}_{1}},{{\mathbf{p}}_{2}},{{\mathbf{p}}_{3}},{{\mathbf{p}}_{4}}$, in which is convenient to make the following replacements:
\begin{equation}
\begin{split}
&{{{\mathbf{p}}}_{1}}+{{{\mathbf{p}}}_{2}}={{{\mathbf{P}}}_{12}},~~~\frac{{{{\mathbf{p}}}_{2}}-{{{\mathbf{p}}}_{1}}}{2}={{{\mathbf{p}}}_{12}},\\
&{{{\mathbf{p}}}_{3}}+{{{\mathbf{p}}}_{4}}={{{\mathbf{P}}}_{34}}, ~~~\frac{{{{\mathbf{p}}}_{4}}-{{{\mathbf{p}}}_{3}}}{2}={{{\mathbf{p}}}_{34}}, 
\label{eq:Pip}	
\end{split}
\end{equation}
The Jacobian of which is equal to 1. Considering this substitution, we will obtain:
\begin{equation}
\begin{split}
& T_{00}^{\left( {\mathbf{R}} \right)}\left( {\mathbf{R}} \right)=\frac{1}{{{\left( 2\pi  \right)}^{3}}}\int{d{{{\mathbf{P}}}_{12}}d{{{\mathbf{P}}}_{34}}}d{{{\mathbf{p}}}_{12}}d{{{\mathbf{p}}}_{34}} \left( \frac{{{\left( {{{\mathbf{P}}}_{34}} \right)}^{2}}}{4m} \right)\\
& \times {{{\hat{\bar{q}}}}^{+}}\left( {{\xi }_{1}},{{{\mathbf{p}}}_{1}}=\frac{1}{2}{{{\mathbf{P}}}_{12}}-{{{\mathbf{p}}}_{12}} \right){{{\hat{q}}}^{+}}\left( {{\xi }_{2}},{{{\mathbf{p}}}_{2}}=\frac{1}{2}{{{\mathbf{P}}}_{12}}+{{{\mathbf{p}}}_{12}} \right) \\ 
& \times {{{\hat{\bar{q}}}}^{-}}\left( {{\xi }_{2}},{{{\mathbf{p}}}_{3}}=\frac{1}{2}{{{\mathbf{P}}}_{34}}-{{{\mathbf{p}}}_{34}} \right){{{\hat{q}}}^{-}}\left( {{\xi }_{2}},{{{\mathbf{p}}}_{4}}=\frac{1}{2}{{{\mathbf{P}}}_{34}}+{{{\mathbf{p}}}_{34}} \right) \\ 
& \times \exp \left( i\left( {{{\mathbf{P}}}_{12}}-{{{\mathbf{P}}}_{34}} \right)\mathbf{R} \right)\delta \left( {{{\mathbf{p}}}_{12}}+{{{\mathbf{p}}}_{34}} \right). \\ 
\end{split}
\label{eq:T00Rimp}
\end{equation}

Here $\delta \left( {{{\mathbf{p}}}_{12}}+{{{\mathbf{p}}}_{34}} \right)$ stands for Dirac $\delta $-function. Due to this function we may perform integration over ${{\mathbf{p}}_{34}}$. With the help of the following denotation
\begin{equation}
{{\mathbf{p}}_{12}}=\mathbf{p},~~~{{\mathbf{p}}_{34}}=-\mathbf{p},
\label{eq:p12_p}
\end{equation}
\eqref{eq:T00Rimp} can be rewritten in the form
\begin{equation}
\begin{split}
T_{00}^{\left( {\mathbf{R}} \right)}\left( {\mathbf{R}} \right)&=\frac{1}{{{\left( 2\pi  \right)}^{3}}}\int{d{{{\mathbf{P}}}_{12}}d{{{\mathbf{P}}}_{34}}}d\mathbf{p}\left( \frac{{{\left( {{{\mathbf{P}}}_{34}} \right)}^{2}}}{4m} \right)\\
& \times {{{\bar{q}}}^{+}}\left( {{\xi }_{1}},{{{\mathbf{p}}}_{1}}=\frac{1}{2}{{{\mathbf{P}}}_{12}}-\mathbf{p} \right)  {{{\hat{q}}}^{+}}\left( {{\xi }_{2}},{{{\mathbf{p}}}_{2}}=\frac{1}{2}{{{\mathbf{P}}}_{12}}+\mathbf{p} \right)\\
& \times {{{\hat{\bar{q}}}}^{-}}\left( {{\xi }_{2}},{{{\mathbf{p}}}_{3}}=\frac{1}{2}{{{\mathbf{P}}}_{34}}+\mathbf{p} \right) {{{\hat{q}}}^{-}}\left( {{\xi }_{2}},{{{\mathbf{p}}}_{4}}=\frac{1}{2}{{{\mathbf{P}}}_{34}}-\mathbf{p} \right)\\
&\times \exp \left( i\left( {{{\mathbf{P}}}_{12}}-{{{\mathbf{P}}}_{34}} \right)\mathbf{R} \right). \\ 
\end{split}
\label{eq:T00Rimp1}
\end{equation}

The expression for $T_{00}^{\left( {\mathbf{r}} \right)}\left( {\mathbf{R}} \right)$ is analogous to $T_{00}^{\left( {\mathbf{R}} \right)}\left( {\mathbf{R}} \right)$, but taking into account that operator ${{\Delta }_{{\mathbf{r}}}}$ provides the other variable differentiation of exponent we get:
\begin{equation}
\begin{split}
T_{00}^{\left( {\mathbf{r}} \right)}\left( {\mathbf{R}} \right)&=\frac{1}{{{\left( 2\pi  \right)}^{3}}}\int{d{{{\mathbf{P}}}_{12}}d{{{\mathbf{P}}}_{34}}}d\mathbf{p}\left( \frac{{{{\mathbf{p}}}^{2}}}{m} \right)\\
& \times {{{\hat{\bar{q}}}}^{+}}\left( {{\xi }_{1}},{{{\mathbf{p}}}_{1}}=\frac{1}{2}{{{\mathbf{P}}}_{12}}-\mathbf{p} \right){{{\hat{q}}}^{+}}\left( {{\xi }_{2}},{{{\mathbf{p}}}_{2}}=\frac{1}{2}{{{\mathbf{P}}}_{12}}+\mathbf{p} \right) \\ 
& \times {{{\hat{\bar{q}}}}^{-}}\left( {{\xi }_{2}},{{{\mathbf{p}}}_{3}}=\frac{1}{2}{{{\mathbf{P}}}_{34}}+\mathbf{p} \right){{{\hat{q}}}^{-}}\left( {{\xi }_{2}},{{{\mathbf{p}}}_{4}}=\frac{1}{2}{{{\mathbf{P}}}_{34}}-\mathbf{p} \right)\\
&\times \exp \left( i\left( {{{\mathbf{P}}}_{12}}-{{{\mathbf{P}}}_{34}} \right)\mathbf{R} \right). \\  
\end{split}
\end{equation}
Considering \eqref{eq:potential} for $T_{00}^{\left( V \right)}\left( {\mathbf{R}} \right)$ we can write
\begin{equation}
\begin{split}
& T_{00}^{\left( V \right)}\left( {\mathbf{R}} \right)=\frac{1}{{{\left( 2\pi  \right)}^{{9}/{2}\;}}}\int{d{{{\mathbf{P}}}_{12}}d{{{\mathbf{P}}}_{34}}}d{{{\mathbf{p}}}_{12}}d{{{\mathbf{p}}}_{34}}d\mathbf{k}  \cdot V ({\mathbf{k}})\\
& \times{{{\hat{\bar{q}}}}^{+}}\left( {{\xi }_{1}},{{{\mathbf{p}}}_{1}}=\frac{1}{2}{{{\mathbf{P}}}_{12}}-{{{\mathbf{p}}}_{12}} \right){{{\hat{q}}}^{+}}\left( {{\xi }_{2}},{{{\mathbf{p}}}_{2}}=\frac{1}{2}{{{\mathbf{P}}}_{12}}+{{{\mathbf{p}}}_{12}} \right) \\ 
& \times {{{\hat{\bar{q}}}}^{-}}\left( {{\xi }_{2}},{{{\mathbf{p}}}_{3}}=\frac{1}{2}{{{\mathbf{P}}}_{34}}-{{{\mathbf{p}}}_{34}} \right){{{\hat{q}}}^{-}}\left( {{\xi }_{2}},{{{\mathbf{p}}}_{4}}=\frac{1}{2}{{{\mathbf{P}}}_{34}}+{{{\mathbf{p}}}_{34}} \right) \\ 
& \times \exp \left( i\left( {{{\mathbf{P}}}_{12}}-{{{\mathbf{P}}}_{34}} \right)\mathbf{R} \right)\delta \left( {{{\mathbf{p}}}_{12}}+\mathbf{k}+{{{\mathbf{p}}}_{34}} \right). \\  
\end{split}
\label{eq:T00Vimp}
\end{equation}

To integrate expression for $T_{00}^{\left( V \right)}\left( {\mathbf{R}}\right)$ over ${\mathbf{p}}_{34}$ we take into account $\delta$-function with the following substitution
\begin{equation}
{{\mathbf{p}}_{12}}=\mathbf{p},~~~{{\mathbf{p}}_{34}}=-\mathbf{p}-\mathbf{k}.
\label{eq:drugajazamena}
\end{equation}
Then
\begin{equation}
\begin{split}
T_{00}^{\left( V \right)}\left( {\mathbf{R}} \right)&=\frac{1}{{{\left( 2\pi  \right)}^{{9}/{2}\;}}}\int{d{{{\mathbf{P}}}_{12}}d{{{\mathbf{P}}}_{34}}}d\mathbf{p}d\mathbf{k} \cdot V ( {\mathbf{k}})\\
&\times {{{\hat{\bar{q}}}}^{+}}\left( {{\xi }_{1}},{{{\mathbf{p}}}_{1}}=\frac{1}{2}{{{\mathbf{P}}}_{12}}-\mathbf{p} \right)\\
&\times {{{\hat{q}}}^{+}}\left( {{\xi }_{2}},{{{\mathbf{p}}}_{2}}=\frac{1}{2}{{{\mathbf{P}}}_{12}}+\mathbf{p} \right)\\ 
&\times {{{\hat{\bar{q}}}}^{-}}\left( {{\xi }_{2}},{{{\mathbf{p}}}_{3}}=\frac{1}{2}{{{\mathbf{P}}}_{34}}+\mathbf{p}+\mathbf{k} \right)\\
&\times {{{\hat{q}}}^{-}}\left( {{\xi }_{2}},{{{\mathbf{p}}}_{4}}=\frac{1}{2}{{{\mathbf{P}}}_{34}}-\mathbf{p}-\mathbf{k} \right)\\
&\times \exp \left( i\left( {{{\mathbf{P}}}_{12}}-{{{\mathbf{P}}}_{34}} \right)\mathbf{R} \right). \\ 
\end{split}
\label{eq:}
\end{equation}
By substituting the expressions for $T_{00}^{\left( {\mathbf{r}} \right)}\left( {\mathbf{R}} \right) $ and $T_{00}^{\left( V \right)}\left( {\mathbf{R}} \right)$ into Eq.\ref{eq:T00RrV}, and taking into account \eqref{eq:oprT00} and \eqref{eq:Hvnut_vnech} we will obtain the expression for the internal Hamiltonian $\hat{H}\left( {\mathbf{r}} \right)$. Along with this, the $\mathbf{R}$ integration appears, which leads again to appearance of $\delta$-function. This helps us to integrate over ${\mathbf{P}}_{34}$ with the following replacement
\begin{equation}
{{\mathbf{P}}_{12}}=\mathbf{P},~~~{{\mathbf{P}}_{34}}=\mathbf{P}.
\label{eq:poklademo}
\end{equation}
As a result one will obtain: 
\begin{equation}
\begin{split}
{{{\hat{H}}}^{\left( {\mathbf{r}} \right)}}&=\int{T_{00}^{\left( {\mathbf{r}} \right)}\left( {\mathbf{R}} \right)d\mathbf{R}}=\int{d\mathbf{P}}d\mathbf{p}\left( \frac{{{{\mathbf{p}}}^{2}}}{m} \right)\\
&\times {{{\hat{\bar{q}}}}^{+}}\left( {{\xi }_{1}},{{{\mathbf{p}}}_{1}}=\frac{1}{2}\mathbf{P}-\mathbf{p} \right)\\
&\times {{{\hat{q}}}^{+}}\left( {{\xi }_{2}},{{{\mathbf{p}}}_{2}}=\frac{1}{2}\mathbf{P}+\mathbf{p} \right)\\
&\times {{{\hat{\bar{q}}}}^{-}}\left( {{\xi }_{2}},{{{\mathbf{p}}}_{3}}=\frac{1}{2}\mathbf{P}+\mathbf{p} \right)\\
&\times {{{\hat{q}}}^{-}}\left( {{\xi }_{2}},{{{\mathbf{p}}}_{4}}=\frac{1}{2}\mathbf{P}-\mathbf{p} \right), \\ 
{{{\hat{H}}}^{\left( V \right)}}&=\int{T_{00}^{\left( V \right)}\left( {\mathbf{R}} \right)d\mathbf{R}}=\frac{1}{{{\left( 2\pi  \right)}^{{3}/{2}\;}}}\int{d\mathbf{P}}d\mathbf{p}d\mathbf{k}\cdot V({\mathbf{k}})\\
&\times {{{\hat{\bar{q}}}}^{+}}\left( {{\xi }_{1}},{{{\mathbf{p}}}_{1}}=\frac{1}{2}\mathbf{P}-\mathbf{p} \right)\\
&\times {{{\hat{q}}}^{+}}\left( {{\xi }_{2}},{{{\mathbf{p}}}_{2}}=\frac{1}{2}\mathbf{P}+\mathbf{p} \right)\\
&\times {{{\hat{\bar{q}}}}^{-}}\left( {{\xi }_{2}},{{{\mathbf{p}}}_{3}}=\frac{1}{2}\mathbf{P}+\mathbf{p}+\mathbf{k} \right)\\
&\times {{{\hat{q}}}^{-}}\left( {{\xi }_{2}},{{{\mathbf{p}}}_{4}}=\frac{1}{2}\mathbf{P}-\mathbf{p}-\mathbf{k} \right). \\  
\end{split}
\label{eq:vnutrH}
\end{equation}

Let us consider the momentum representation of generator \eqref{eq:oprM03}. For $\hat{M}_{03}^{\left( {\mathbf{R}} \right)}$ we have
\begin{equation}
\begin{split}
\hat{M}_{03}^{\left( {\mathbf{R}} \right)}&=\frac{1}{{{\left( 2\pi  \right)}^{3}}}\int{d{{{\mathbf{P}}}_{12}}d{{{\mathbf{P}}}_{34}}}d\mathbf{p}\left( \frac{{{\left( {{{\mathbf{P}}}_{34}} \right)}^{2}}}{4m} \right) \\ 
& \times{{{\bar{q}}}^{+}}\left( {{\xi }_{1}},{{{\mathbf{p}}}_{1}}=\frac{1}{2}{{{\mathbf{P}}}_{12}}-\mathbf{p} \right){{{\hat{q}}}^{+}}\left( {{\xi }_{2}},{{{\mathbf{p}}}_{2}}=\frac{1}{2}{{{\mathbf{P}}}_{12}}+\mathbf{p} \right) \\ 
& \times {{{\hat{\bar{q}}}}^{-}}\left( {{\xi }_{2}},{{{\mathbf{p}}}_{3}}=\frac{1}{2}{{{\mathbf{P}}}_{34}}+\mathbf{p} \right){{{\hat{q}}}^{-}}\left( {{\xi }_{2}},{{{\mathbf{p}}}_{4}}=\frac{1}{2}{{{\mathbf{P}}}_{34}}-\mathbf{p} \right)\\
&\times \int{{{R}_{3}}\exp \left( i\left( {{{\mathbf{P}}}_{12}}-{{{\mathbf{P}}}_{34}} \right)\mathbf{R} \right)d\mathbf{R}}. \\ 
\end{split}
\label{eq:M03Rimp}
\end{equation}

Integration variable ${{\mathbf{P}}_{12}}$ will be represented as  $\mathbf{P}$. The ${{\mathbf{P}}_{34}}$ will be replaced by 
\begin{equation}
{{\mathbf{P}}_{34}}=\mathbf{P}-\pmb{\varepsilon },
\label{eq:eps}
\end{equation}
where $\pmb{\varepsilon}$ is a new variable of integration.

After this transformations the $\hat{M}_{03}^{\left( {\mathbf{R}} \right)}$ can be written as:
\begin{equation}
\begin{split}
\hat{M}_{03}^{\left( {\mathbf{R}} \right)}&=-i\int{d\mathbf{P}d\pmb{\varepsilon }}d\mathbf{p}\left( \frac{\partial \delta \left( {\pmb{2\varepsilon }} \right)}{\partial {{\varepsilon }_{3}}} \right)\left( \frac{{{\left( \mathbf{P}-\pmb{\varepsilon } \right)}^{2}}}{4m} \right)\\
& \times {{{\bar{q}}}^{+}}\left( {{\xi }_{1}},{{{\mathbf{p}}}_{1}}=\frac{1}{2}\mathbf{P}-\mathbf{p} \right)\\
&\times {{{\hat{q}}}^{+}}\left( {{\xi }_{2}},{{{\mathbf{p}}}_{2}}=\frac{1}{2}\mathbf{P}+\mathbf{p} \right)\\
& \times {{{\hat{\bar{q}}}}^{-}}\left( {{\xi }_{2}},{{{\mathbf{p}}}_{3}}=\frac{1}{2}\left( \mathbf{P}-\pmb{\varepsilon } \right)+\mathbf{p} \right)\\
&\times {{{\hat{q}}}^{-}}\left( {{\xi }_{2}},{{{\mathbf{p}}}_{4}}=\frac{1}{2}\left( \mathbf{P}-\pmb{\varepsilon } \right)-\mathbf{p} \right). 
\end{split}
\label{eq:M03Rimp1}
\end{equation}
Analogously,
\begin{equation}
\begin{split}
\hat{M}_{03}^{\left( {\mathbf{r}} \right)}&=-i\int{d\mathbf{P}d\pmb{\varepsilon }}d\mathbf{p}\left( \frac{\partial \delta \left( {\pmb{2\varepsilon }} \right)}{\partial {{\varepsilon }_{3}}} \right)\left( \frac{{{{\mathbf{p}}}^{2}}}{m} \right)\\
& \times {{{\bar{q}}}^{+}}\left( {{\xi }_{1}},{{{\mathbf{p}}}_{1}}=\frac{1}{2}\mathbf{P}-\mathbf{p} \right)\\
&\times {{{\hat{q}}}^{+}}\left( {{\xi }_{2}},{{{\mathbf{p}}}_{2}}=\frac{1}{2}\mathbf{P}+\mathbf{p} \right) \\
&\times {{{\hat{\bar{q}}}}^{-}}\left( {{\xi }_{2}},{{{\mathbf{p}}}_{3}}=\frac{1}{2}\left(\mathbf{P}-\pmb{\varepsilon } \right)+\mathbf{p} \right) \\
&\times {{{\hat{q}}}^{-}}\left( {{\xi }_{2}},{{{\mathbf{p}}}_{4}}=\frac{1}{2}\left(\mathbf{P}-\pmb{\varepsilon } \right)-\mathbf{p} \right). \\ 
\hat{M}_{03}^{\left( V \right)}&=\frac{-i}{{{\left( 2\pi  \right)}^{{3}/{2}\;}}}\int{d\mathbf{P}d\pmb{\varepsilon }}d\mathbf{p}d\mathbf{k}\left(\frac{\partial \delta \left( {\pmb{2\varepsilon }} \right)}{\partial {{\varepsilon }_{3}}}\right)V\left( {\mathbf{k}} \right)\\
& \times {{{\hat{\bar{q}}}}^{+}}\left( {{\xi }_{1}},{{{\mathbf{p}}}_{1}}=\frac{1}{2}\mathbf{P}-\mathbf{p} \right){{{\hat{q}}}^{+}}\left( {{\xi }_{2}},{{{\mathbf{p}}}_{2}}=\frac{1}{2}\mathbf{P}+\mathbf{p} \right)\\
& \times {{{\hat{\bar{q}}}}^{-}}\left( {{\xi }_{2}},{{{\mathbf{p}}}_{3}}=\frac{1}{2}\left( \mathbf{P}-\pmb{\varepsilon } \right)+\mathbf{p}+\mathbf{k} \right)\\
&\times {{{\hat{q}}}^{-}}\left( {{\xi }_{2}},{{{\mathbf{p}}}_{4}}=\frac{1}{2}\left( \mathbf{P}-\pmb{\varepsilon } \right)-\mathbf{p}-\mathbf{k} \right).
\end{split}
\label{eq:M03Rimp1mal}
\end{equation}

Knowing the expressions for the generator ${{\hat{M}}_{03}}$ and the internal Hamiltonian, we can calculate their commutator. Since each of those operators consists of several terms, let us consider commutators between those terms. While calculating the commutators, it is convenient firstly to convert the products of operators arranged in that or another order into the normal form with the help of Wick's theorem. 

Above, we have already used the fact that all operators are considered in a subspace of the Fock space, whose states include one quark and one antiquark. Therefore, the operators containing, in the normal form, two or more quark/antiquark creation or annihilation operators will have zero matrix elements for all basis elements of this subspace, so that such operators can be dropped. Taking into account that each of the operators concerned contains one quark and one antiquark operators, their product will include two quark and two antiquark operators, thus containing “redundant” operators. Applying Wick's theorem to this product, we obtain that only those terms will have nonzero matrix elements in the space concerned, in which the “redundant” operators are paired.

Let us analyze the product $\hat{H}^{\left( {\mathbf{r}} \right)}\hat{M}_{03}^{\left(\mathbf{R} \right)}$. After converting it to the normal form, dropping the terms with zero matrix elements, this product can be written in the form
\begin{equation}
\begin{split}
& {{{\hat{H}}}^{\left( {\mathbf{r}} \right)}}\hat{M}_{03}^{\left( {\mathbf{R}} \right)}=-i\int{d\mathbf{P}}d\mathbf{p}d{\mathbf{P}}'d{\mathbf{p}}'d\pmb{\varepsilon }\left( \frac{{{{\mathbf{p}}}^{2}}}{m} \right)\left( \frac{{{\left( {\mathbf{P}}'-\pmb{\varepsilon } \right)}^{2}}}{4m} \right)\left( \frac{\partial \delta \left( {\pmb{2\varepsilon }} \right)}{\partial {{\varepsilon }_{3}}} \right) \\ 
&\times \delta \left( \left( \frac{1}{2}\mathbf{P}+\mathbf{p} \right)-\left( \frac{1}{2}{\mathbf{P}}'+{\mathbf{p}}' \right) \right)\\
&\times \delta \left( \left( \frac{1}{2}\mathbf{P}-\mathbf{p} \right)-\left( \frac{1}{2}{\mathbf{P}}'-{\mathbf{p}}' \right) \right) \\ 
&\times {{{\hat{\bar{q}}}}^{+}}\left( {{\xi }_{1}},{{{\mathbf{p}}}_{1}}=\frac{1}{2}\mathbf{P}-\mathbf{p} \right)\\
&\times {{{\hat{q}}}^{+}}\left( {{\xi }_{2}},{{{\mathbf{p}}}_{2}}=\frac{1}{2}\mathbf{P}+\mathbf{p} \right) \\ 
&\times {{{\hat{\bar{q}}}}^{-}}\left( {{\xi }_{2}},{{{\mathbf{p}}}_{3}}=\frac{1}{2}\left( {\mathbf{P}}'-\pmb{\varepsilon } \right)+{\mathbf{p}}' \right)\\
&\times {{{\hat{q}}}^{-}}\left( {{\xi }_{1}},{{{\mathbf{p}}}_{4}}=\frac{1}{2}\left( {\mathbf{P}}'-\pmb{\varepsilon } \right)-{\mathbf{p}}' \right). \\ 
\end{split}
\label{eq:Rr}
\end{equation}    

Here we have already considered the subsum on the internal indexes $\xi$ with the Kronecker $\delta$-symbols that appears in the paired operators. Note, that during the integration we returned back to the old variables
\begin{equation}
\begin{split}
& {{{\mathbf{p}}}_{1}}=\frac{1}{2}\mathbf{P}-\mathbf{p},~~~~{{{\mathbf{p}}}_{2}}=\frac{1}{2}\mathbf{P}+\mathbf{p}, \\ 
& {{{\mathbf{p}}}_{3}}=\frac{1}{2}{\mathbf{P}}'+{\mathbf{p}}',~~~~{{{\mathbf{p}}}_{4}}=\frac{1}{2}{\mathbf{P}}'-{\mathbf{p}}'. \\ 
\end{split}
\label{eq:starzmin}
\end{equation}

Afterwards, due to $\delta $-functions one may perform integration of the 
${{\mathbf{p}}_{3}}$ and ${{\mathbf{p}}_{4}}$. As a result, obtain
\begin{equation}
\begin{split}
{{{\hat{H}}}^{\left( {\mathbf{r}} \right)}}\hat{M}_{03}^{\left( {\mathbf{R}} \right)}&=-i\int{d{{{\mathbf{p}}}_{1}}}d{{{\mathbf{p}}}_{2}}d\pmb{\varepsilon }\\
&\times\left( \frac{\partial \delta \left( {\pmb{2\varepsilon }} \right)}{\partial {{\varepsilon }_{3}}} \right)\left( \frac{{{\left( {{{\mathbf{p}}}_{2}}-{{{\mathbf{p}}}_{1}} \right)}^{2}}}{4m} \right)\left( \frac{{{\left( {{{\mathbf{p}}}_{1}}+{{{\mathbf{p}}}_{2}}-\pmb{\varepsilon } \right)}^{2}}}{4m} \right) \\ 
&\times {{{\hat{\bar{q}}}}^{+}}\left( {{\xi }_{1}},{{{\mathbf{p}}}_{1}} \right){{{\hat{q}}}^{+}}\left( {{\xi }_{2}},{{{\mathbf{p}}}_{2}} \right)\\
&\times {{{\hat{\bar{q}}}}^{-}}\left( {{\xi }_{2}},{{{\mathbf{p}}}_{2}}-\frac{1}{2}\pmb{\varepsilon } \right) {{{\hat{q}}}^{-}}\left( {{\xi }_{1}},{{{\mathbf{p}}}_{1}}-\frac{1}{2}\pmb{\varepsilon } \right). \\ 
\end{split}
\label{eq:Rrupr2}
\end{equation} 
For the product of the same operators, but in the inverse order, after similar transformations, we obtain
\begin{equation}
\begin{split}
\hat{M}_{03}^{\left( {\mathbf{R}} \right)}{{{\hat{H}}}^{\left( {\mathbf{r}} \right)}}&=-i\int{d{{{\mathbf{p}}}_{1}}d{{{\mathbf{p}}}_{2}}}d{{{\mathbf{p}}}_{3}}d{{{\mathbf{p}}}_{4}}d\pmb{\varepsilon }\\
&\times \left( \frac{{{\left( {{{\mathbf{p}}}_{3}}-{{{\mathbf{p}}}_{4}} \right)}^{2}}}{4m} \right)\left( \frac{{{\left( {{{\mathbf{p}}}_{1}}+{{{\mathbf{p}}}_{2}}-\pmb{\varepsilon } \right)}^{2}}}{4m} \right)\left( \frac{\partial \delta \left( {\pmb{2\varepsilon }} \right)}{\partial {{\varepsilon }_{3}}} \right) \\ 
& \times \delta \left( \left( {{{\vec{p}}}_{2}}-\frac{1}{2}\pmb{\varepsilon } \right)-{{{\mathbf{p}}}_{3}} \right)\delta \left( \left( {{{\mathbf{p}}}_{1}}-\frac{1}{2}\pmb{\varepsilon } \right)-{{{\mathbf{p}}}_{4}} \right)\\
&\times {{{\bar{q}}}^{+}}\left( {{\xi }_{1}},{{{\mathbf{p}}}_{1}} \right){{{\hat{q}}}^{+}}\left( {{\xi }_{2}},{{{\mathbf{p}}}_{2}} \right)\\
&\times {{{\hat{\bar{q}}}}^{-}}\left( {{\xi }_{2}},{{{\mathbf{p}}}_{3}} \right){{{\hat{q}}}^{-}}\left( {{\xi }_{1}},{{{\mathbf{p}}}_{4}} \right). \\  
\end{split}
\label{eq:Rrobrporadok}
\end{equation}

Carrying out the integration over the components ${{\mathbf{p}}_{3}}$ and ${{\mathbf{p}}_{4}}$, we obtain a result that is identical to expression \eqref{eq:Rrupr2}. Hence, one may conclude that operators $\hat{M}_{03}^{\left( {\mathbf{R}} \right)} $ and ${{\hat{H}}^{\left( {\mathbf{r}} \right)}}$ commute with each other.

Now, let us consider the commutator $\left[ \hat{M}_{03}^{\left( {\mathbf{R}} \right)},{{{\hat{H}}}^{\left( V \right)}} \right]$. After converting it to the normal form, and with the use of the transformation considered above (returning to the ``old''), this product can be written in the form
\begin{equation}
\begin{split}
\hat{M}_{03}^{\left( {\mathbf{R}} \right)}{{{\hat{H}}}^{\left( V \right)}}&=-i\frac{1}{{{\left( 2\pi  \right)}^{{3}/{2}\;}}}\int{d{{{\mathbf{p}}}_{1}}d{{{\mathbf{p}}}_{2}}}d\mathbf{k}d\pmb{\varepsilon }\\
&\times \left( \frac{\partial \delta \left( {2\pmb{\varepsilon }} \right)}{\partial {{\varepsilon }_{3}}} \right)\left( \frac{{{\left( {{{\mathbf{p}}}_{2}}+{{{\mathbf{p}}}_{1}}-\pmb{\varepsilon } \right)}^{2}}}{4m} \right)V\left( {\mathbf{k}} \right) \\ 
& \times {{{\bar{q}}}^{+}}\left( {{\xi }_{1}},{{{\mathbf{p}}}_{1}} \right){{{\hat{q}}}^{+}}\left( {{\xi }_{2}},{{{\mathbf{p}}}_{2}} \right)\\
&\times {{{\hat{\bar{q}}}}^{-}}\left( {{\xi }_{2}},{{{\mathbf{p}}}_{3}}=\left( {{{\mathbf{p}}}_{2}}-\frac{1}{2}\pmb{\varepsilon } \right)+\mathbf{k} \right)\\
&\times {{{\hat{q}}}^{-}}\left( {{\xi }_{1}},{{{\mathbf{p}}}_{4}}=\left( {{{\mathbf{p}}}_{1}}-\frac{1}{2}\pmb{\varepsilon } \right)-\mathbf{k} \right). \\ 
\end{split}
\label{eq:MRHVpra}
\end{equation}

Calculating the product of these operators in inverse order ${{\hat{H}}^{\left( V \right)}}\hat{M}_{03}^{\left( {\mathbf{R}} \right)}$ is similar, and leads to the expression which coincides with the right-hand side of \eqref{eq:MRHVpra}. Therefore, the operator $\hat{M}_{03}^{\left( {\mathbf{R}} \right)}$ commutes with each one of the two terms of internal Hamiltonian, see \eqref{eq:vnutrH1}, and thus commutes with the Hamiltonian of the bound quarks system.

Let us now calculate the commutators of the operator $\hat{M}_{03}^{\left( {\mathbf{r}} \right)}$ with the summands of internal Hamiltonian. With the help of transformations considered above a product ${\hat{H}}^{\left( \mathbf{r} \right)} \hat{M}_{03}^{\left( {\mathbf{r}} \right)}$ is reduced to the form:
\begin{equation}
\begin{split}
{{{\hat{H}}}^{\left( {\mathbf{r}} \right)}}\hat{M}_{03}^{\left( {\mathbf{r}} \right)}&=-i\int{d{{{\mathbf{p}}}_{1}}d{{{\mathbf{p}}}_{2}}d{{{\mathbf{p}}}_{3}}d{{{\mathbf{p}}}_{4}}d\pmb{\varepsilon }}\\
&\times \left( \frac{\partial \delta \left( {2\pmb{\varepsilon }} \right)}{\partial {{\varepsilon }_{3}}} \right)\left( \frac{{{\left( {{{\mathbf{p}}}_{2}}-{{{\mathbf{p}}}_{1}} \right)}^{2}}}{4m} \right)\left( \frac{{{\left( {{{\mathbf{p}}}_{2}}-{{{\mathbf{p}}}_{1}} \right)}^{2}}}{4m} \right) \\ 
&\times {{{\hat{\bar{q}}}}^{+}}\left( {{\xi }_{1}},{{{\mathbf{p}}}_{1}} \right){{{\hat{q}}}^{+}}\left( {{\xi }_{2}},{{{\mathbf{p}}}_{2}} \right)\\
&\times {{{\hat{\bar{q}}}}^{-}}\left( {{\xi }_{2}},{{{\mathbf{p}}}_{3}}={{{\mathbf{p}}}_{2}}-\frac{1}{2}\pmb{\varepsilon } \right){{{\hat{q}}}^{-}}\left( {{\xi }_{1}},{{{\mathbf{p}}}_{4}}={{{\mathbf{p}}}_{1}}-\frac{1}{2}\pmb{\varepsilon } \right). \\  
\end{split}
\label{eq:MrHrpra}
\end{equation}

The same result is obtained if one construct the normal form from these operators product but in inverse order.  Thus $\hat{M}_{03}^{\left( \mathbf{r} \right)}$ and $\hat{H}^{\left( {\mathbf{r}} \right)}$ commutes with each other. 

Calculation of the product ${{\hat{H}}^{\left( V \right)}}\hat{M}_{03}^{\left( {\mathbf{r}} \right)}$ gives us
\begin{equation}
\begin{split}
{{{\hat{H}}}^{\left( V \right)}}\hat{M}_{03}^{\left( {\mathbf{r}} \right)}&=\frac{-i}{{{\left( 2\pi  \right)}^{{3}/{2}\;}}}\int{d{{{\mathbf{p}}}_{1}}d{{{\mathbf{p}}}_{2}}d\pmb{\varepsilon }}d\mathbf{k}\\
&\times \left( \frac{\partial \delta \left( {2\pmb{\varepsilon }} \right)}{\partial {{\varepsilon }_{3}}} \right)V\left( {\mathbf{k}} \right)\left( \frac{{{\left( {{{\mathbf{p}}}_{2}}-{{{\mathbf{p}}}_{1}}+2\mathbf{k} \right)}^{2}}}{4m} \right) \\ 
&\times {{{\hat{\bar{q}}}}^{+}}\left( {{\xi }_{1}},{{{\mathbf{p}}}_{1}} \right){{{\hat{q}}}^{+}}\left( {{\xi }_{2}},{{{\mathbf{p}}}_{2}} \right)\\
&\times {{{\hat{\bar{q}}}}^{-}}\left( {{\xi }_{2}},{{{\mathbf{p}}}_{3}}={{{\mathbf{p}}}_{2}}+\mathbf{k}-\frac{1}{2}\pmb{\varepsilon } \right)\\
&\times {{{\hat{q}}}^{-}}\left( {{\xi }_{1}},{{{\mathbf{p}}}_{4}}={{{\mathbf{p}}}_{1}}-\mathbf{k}-\frac{1}{2}\pmb{\varepsilon } \right). \\ 
\end{split}
\label{eq:HVMrpram2}
\end{equation}

The calculated product of these operators in inverse order $\hat{M}_{03}^{\left( {\mathbf{r}} \right)}{{\hat{H}}^{\left( V \right)}}$ gives the result that does not coincide with \eqref{eq:HVMrpram2}, therefore operators $\hat{M}_{03}^{\left( {\mathbf{r}} \right)}$ and ${\hat{H}}^{\left( V \right)}$ do not commute with each other:
\begin{equation}
\begin{split}
\hat{M}_{03}^{\left( {\mathbf{r}} \right)}{{{\hat{H}}}^{\left( V \right)}}&=\frac{-i}{{{\left( 2\pi  \right)}^{{3}/{2}\;}}}\int{d{{{\mathbf{p}}}_{1}}d{{{\mathbf{p}}}_{2}}d\pmb{\varepsilon }}d\mathbf{k}\\
&\times V\left( {\mathbf{k}} \right)\left( \frac{\partial \delta \left( {2\pmb{\varepsilon }} \right)}{\partial {{\varepsilon }_{3}}} \right)\left( \frac{{{\left( {{{\mathbf{p}}}_{2}}-{{{\mathbf{p}}}_{1}} \right)}^{2}}}{4m} \right) \\ 
&\times {{{\bar{q}}}^{+}}\left( {{\xi }_{1}},{{{\mathbf{p}}}_{1}} \right){{{\hat{q}}}^{+}}\left( {{\xi }_{2}},{{{\mathbf{p}}}_{2}} \right)\\
&\times {{{\hat{\bar{q}}}}^{-}}\left( {{\xi }_{2}},{{{\mathbf{p}}}_{3}}={{{\mathbf{p}}}_{2}}+\mathbf{k}-\frac{1}{2}\pmb{\varepsilon } \right)\\
&\times {{{\hat{q}}}^{-}}\left( {{\xi }_{2}},{{{\mathbf{p}}}_{4}}={{{\mathbf{p}}}_{1}}-\mathbf{k}-\frac{1}{2}\pmb{\varepsilon } \right). \\ 
\end{split}
\label{eq:HVMrobr2}
\end{equation}
But
\begin{equation}
\begin{split}
\hat{M}_{03}^{\left( V \right)}{{{\hat{H}}}^{\left( {\mathbf{r}} \right)}}&=\frac{-i}{{{\left( 2\pi  \right)}^{{3}/{2}\;}}}\int{d{{{\mathbf{p}}}_{1}}d{{{\mathbf{p}}}_{2}}d{{{\mathbf{p}}}_{3}}d{{{\mathbf{p}}}_{4}}d\pmb{\varepsilon }}d\mathbf{k}\\
&\times \left(\frac{\partial \delta \left( {2\pmb{\varepsilon }} \right)}{\partial {{\varepsilon }_{3}}} \right) V\left( {\mathbf{k}} \right)\left( \frac{{{\left( {{{\mathbf{p}}}_{2}}-{{{\mathbf{p}}}_{1}}+2\mathbf{k} \right)}^{2}}}{4m} \right) \\ 
& \times {{{\hat{\bar{q}}}}^{+}}\left( {{\xi }_{1}},{{{\mathbf{p}}}_{1}} \right){{{\hat{q}}}^{+}}\left( {{\xi }_{2}},{{{\mathbf{p}}}_{2}} \right)\\
&\times {{{\hat{\bar{q}}}}^{-}}\left( {{\xi }_{2}},{{{\mathbf{p}}}_{3}}={{{\mathbf{p}}}_{2}}+\mathbf{k}-\frac{1}{2}\pmb{\varepsilon } \right)\\
&\times {{{\hat{q}}}^{-}}\left( {{\xi }_{1}},{{{\mathbf{p}}}_{4}}={{{\mathbf{p}}}_{1}}-\mathbf{k}-\frac{1}{2}\pmb{\varepsilon } \right). \\ 
\end{split}
\label{eq:MVHrpram2}
\end{equation}
and 
\begin{equation}
\begin{split}
{{{\hat{H}}}^{\left( {\mathbf{r}} \right)}}\hat{M}_{03}^{\left( V \right)}&=\frac{-i}{{{\left( 2\pi  \right)}^{{3}/{2}\;}}}\int{d{{{\mathbf{p}}}_{1}}d{{{\mathbf{p}}}_{2}}d{{{\mathbf{p}}}_{3}}d{{{\mathbf{p}}}_{4}}}d\pmb{\varepsilon }d\mathbf{k}\\
&\times \frac{\partial \delta \left( {2\pmb{\varepsilon }} \right)}{\partial {{\varepsilon }_{3}}}V\left( {\mathbf{k}} \right)\left( \frac{{{\left( {{{\mathbf{p}}}_{2}}-{{{\mathbf{p}}}_{1}} \right)}^{2}}}{m} \right) \\ 
& \times {{{\hat{\bar{q}}}}^{+}}\left( {{\xi }_{1}},{{{\mathbf{p}}}_{1}} \right){{{\hat{q}}}^{+}}\left( {{\xi }_{2}},{{{\mathbf{p}}}_{2}} \right)\\
&\times {{{\hat{\bar{q}}}}^{-}}\left( {{\xi }_{2}},{{{\mathbf{p}}}_{3}}={{{\mathbf{p}}}_{2}}+\mathbf{k}-\frac{1}{2}\pmb{\varepsilon } \right)\\
&\times {{{\hat{q}}}^{-}}\left( {{\xi }_{1}},{{{\mathbf{p}}}_{4}}={{{\mathbf{p}}}_{1}}-\mathbf{k}-\frac{1}{2}\pmb{\varepsilon } \right). \\ 
\end{split}
\label{eq:MVHobr2}
\end{equation}

Taking into account that the expressions for ${{\hat{H}}^{\left( V \right)}}\hat{M}_{03}^{\left( {\mathbf{r}} \right)}$ and $\hat{M}_{03}^{\left( V \right)}{{\hat{H}}^{\left( {\mathbf{r}} \right)}}$ [Eqs.(\ref{eq:HVMrpram2}) and (\ref{eq:MVHrpram2}), respectively] enter the general expression for the commutator $\left[ {{{\hat{H}}}^{\left( \mathbf{r},V \right)}},{{{\hat{M}}}_{03}} \right]$ with opposite signs (the same can be said about Eqs.(\ref{eq:HVMrobr2}) and (\ref{eq:MVHobr2})), we arrive at a conclusion that
\begin{equation}
\left[ {{{\hat{H}}}^{\left( {\mathbf{r}} \right)}},\hat{M}_{03}^{\left( V \right)} \right]+\left[ {{{\hat{H}}}^{\left( V \right)}},\hat{M}_{03}^{\left( {\mathbf{r}} \right)} \right]=0.
\label{eq:CommutatorVrrV}
\end{equation}
Finally, computation of products $\hat{M}_{03}^{\left( V \right)}{{\hat{H}}^{\left( V \right)}} $ and ${{\hat{H}}^{\left( V \right)}}\hat{M}_{03}^{\left( V \right)}$ give us the same result:

\begin{equation}
\begin{split}
\hat{M}_{03}^{\left( V \right)}{{{\hat{H}}}^{\left( V \right)}}&={{{\hat{H}}}^{\left( V \right)}}\hat{M}_{03}^{\left( V \right)}\\
&=\frac{-i}{{{\left( 2\pi  \right)}^{3}}}\int{d{{{\mathbf{p}}}_{1}}d{{{\mathbf{p}}}_{2}}}d\pmb{\varepsilon }d{\mathbf{p}}'d{\mathbf{k}}' \left( \frac{\partial \delta \left( {2\pmb{\varepsilon }} \right)}{\partial {{\varepsilon }_{3}}} \right) V\left( {\mathbf{k}} \right)V\left( {{\mathbf{k}}'} \right) \\ 
&\times {{{\hat{\bar{q}}}}^{+}}\left( {{\xi }_{1}},{{{\mathbf{p}}}_{1}} \right) {{{\hat{q}}}^{+}}\left( {{\xi }_{2}},{{{\mathbf{p}}}_{2}} \right)\\
&\times {{{\hat{\bar{q}}}}^{-}}\left( {{\xi }_{2}},{{{\mathbf{p}}}_{3}}={{{\mathbf{p}}}_{2}}+{\mathbf{k}}'+\mathbf{k}-\frac{1}{2}\pmb{\varepsilon } \right)\\
&\times {{{\hat{q}}}^{-}}\left( {{\xi }_{1}},{{{\mathbf{p}}}_{4}}={{{\mathbf{p}}}_{1}}-{\mathbf{k}}'-\mathbf{k}-\frac{1}{2}\pmb{\varepsilon } \right). \\ 
\end{split}
\label{eq:MVHVpramobr}
\end{equation}

Therefore, if we decompose the commutator $\left[ {{{\hat{H}}}^{\left( \mathbf{r},V \right)}},{{{\hat{M}}}_{03}} \right]$ on the terms that corresponds to the summands of internal Hamiltonian and generator, then the sum of all these summands is equal to zero, hence
\begin{equation}
\left[ {{{\hat{M}}}_{03}},{{{\hat{H}}}^{\left( \mathbf{r},V \right)}} \right]=0,
\label{eq:polnijcommutator}
\end{equation}
which was to be proved.




\bibliography{UkrJPhys_references}

\begin{thebibliography}{10}

\bibitem{ujp}
Sharf I, et~al. (2011) {Description of hadron inelastic scattering by the
  Laplace method and new mechanisms of cross-section growth}.
\newblock {\em Ukr.J.Phys.} 56:1151--1164.

\bibitem{cej}
Sharf I, et~al. (2012) {On the Role of Longitudinal Momenta in High Energy
  Hadron-Hadron Scattering}.
\newblock {\em Central Eur.J.Phys.} 10:858--887.
\newblock \textcolor{blue}{[arXiv:1110.4945/hep-ph]}.

\bibitem{Sharph:2015eka}
Sharph IV, , et~al. (2015) {The new method of interference contributions
  accounting for inelastic scattering diagrams}.
\newblock \textcolor{blue}{[arXiv:1509.04329/hep-ph]}.

\bibitem{Sharf:2012vy}
Sharf IV, et~al. (2012) {Gluon Loops in the Inelastic Processes in QCD}.
\newblock \textcolor{blue}{[arXiv:1210.3490/hep-ph]}.

\bibitem{Volkotrub:2016}
Volkotrub Y, , et~al. (2016) {Multi-particle field operators in quantum field
  theory}.
\newblock \textcolor{blue}{[arXiv:1510.01937/physics.gen-ph]}.

\bibitem{Chudak:2016}
Chudak N, , et~al. (2016) {Multi-Particle Quantum Fields}.
\newblock {\em AIS Physics Journal} 2(3):181--195.

\bibitem{Feynman:1972}
{Feynman, Richard F.} ({1972}) {\em {Photon-Hadron Interactions}}.
\newblock ({W.A. Benjamin, Inc. Publisher}, {Boston}), p. {298}.

\bibitem{Diehl:2011yj}
Diehl M, Ostermeier D, Schafer A (2012) {Elements of a theory for multiparton
  interactions in QCD}.
\newblock {\em JHEP} 1203:089.
\newblock \textcolor{blue}{[arXiv:1111.0910/hep-ph]}.

\bibitem{Strikman:2011zz}
Strikman M (2011) {Transverse structure of the nucleon and multiparton
  interactions}.
\newblock {\em Prog.Theor.Phys.Suppl.} 187:289--296.
\newblock "High Energy Strong Interactions 2010: Parton Distributions and Dense
  QCD Matter -- Proceedings of the YIPQS International Workshop".

\bibitem{Kobushkin:1977}
Kobushkin AP, Shelest VP ({1977}) Relativistic equations for bound quark
  systems.
\newblock {\em {Teor. Mat. Fiz.}} 31:156.
\newblock {[in Russian]}.

\bibitem{FAUSTOV1973176}
Faustov R (1973) Relativistic wavefunction and form factors of the bound
  system.
\newblock {\em Annals of Physics} 78(1):176 -- 189.

\bibitem{PhysRev.84.1232}
Salpeter EE, Bethe HA (1951) A relativistic equation for bound-state problems.
\newblock {\em Phys. Rev.} 84(6):1232--1242.

\bibitem{Logunov1963}
{Logunov, A. A. and Tavkhelidze, A. N.} (1963) {Quasi-optical approach in
  quantum field theory}.
\newblock {\em {Il Nuovo Cimento (1955-1965)}} 29(2):380--399.

\bibitem{Brodsky:1997de}
Brodsky SJ, Pauli HC, Pinsky SS (1998) {Quantum chromodynamics and other field
  theories on the light cone}.
\newblock {\em Phys.Rept.} 301:299--486.
\newblock \textcolor{blue}{[arXiv:9705477/hep-ph]}.

\bibitem{Heinzl:1998kz}
Heinzl T (1998) Ph.D. thesis (Regensburg U.).
\newblock \textcolor{blue}{[arXiv:9812190/hep-th]}.

\bibitem{Terentev:1976}
Terent’ev M (1976) On the structure of wave functions of mesons as bound
  quark states.
\newblock {\em Yad. Fiz.} 24:207.

\bibitem{RevModPhys.21.392}
Dirac PAM (1949) Forms of relativistic dynamics.
\newblock {\em Rev. Mod. Phys.} 21(3):392--399.

\bibitem{Bogolubov:1959}
Bogolyubov N, Shirkov D (1959) {Introduction To The Theory Of Quantized
  Fields}.
\newblock {\em Intersci.Monogr.Phys.Astron.} 3:1--720.

\bibitem{Berezin:1966}
{Berezin, F.A.} ({1966}) {\em {The Method of Second Quantization (Monographs
  and Textbooks in Pure and Applied Physics, Vol. 24)}}.
\newblock ({Academic Prfess}, {New York and London}), {1st} edition, p. {228}.
\newblock {Translated by Nobumichi Mugibayashi and Alan Jeffrey}.

\bibitem{Gelfand:1964}
Gel'fand IM, Shilov GE (1964-1968) {\em Generalized Functions}.
\newblock ({Academic Press}, {New York}), {1st} edition, p. {423}.

\bibitem{Karmanov:1988}
Karmanov V (1988) Relativistic composite systems in the light front dynamics.
\newblock {\em Fiz. Elem. Chast. At. Yadra} 19:525.
\newblock {[in Russian]}.

\bibitem{Raider}
Ryder L (1985) {\em Quantum Field Theory}.
\newblock ({Cambridge University Press}, {Cambridge}), p. {501}.

\bibitem{XiandongJi:2004}
Xiandong J (2004) {Generalized Parton Distributions}.
\newblock {\em Annual Review of Nuclear and Particle Science} 54(1):413--450.

\bibitem{RevModPhys.84.1231}
Alexandrou C, Papanicolas CN, Vanderhaeghen M (2012) {Colloquium: The Shape of
  Hadrons}.
\newblock {\em Rev. Mod. Phys.} 84(3):1231--1251.
\newblock \textcolor{blue}{[arXiv:1201.4511/hep-ph]}.

\bibitem{Dremin:2013}
Dremin IM (2013) Elastic scattering of hadrons.
\newblock {\em Uspekhi Fizicheskikh Nauk} 183(1):3--32.
\newblock \textcolor{blue}{[arXiv:1206.5474/hep-ph]}.

\bibitem{Conceicao201258}
Conceicao R, de~Deus JD, Pimenta M (2012) Proton–proton cross-sections: The
  interplay between density and radius.
\newblock {\em Nuclear Physics A} 888:58 -- 66.
\newblock \textcolor{blue}{[arXiv:1107.0912/hep-ph]}.

\bibitem{Sharph:2014nza}
Sharph IV, , et~al. (2014) {The state of nonrelativistic quantum system in a
  relativistic reference frame}.
\newblock \textcolor{blue}{[arXiv:1403.3114/hep-ph]}.

\end{thebibliography}

\end{document}